\documentclass[
  twocolumn,
  prb,
  showpacs,
  amsmath,
  amssymb,
  superscriptaddress
]{revtex4}

\usepackage{bm}
\usepackage{graphicx}
\usepackage{epstopdf}

\usepackage{hyperref}
\begin{document}

\title{Pulse propagation  in interacting one dimensional  Bose liquid}

\author{A.\ D.\ Sarishvili}
\affiliation{Department of Physics, Bar Ilan University, Ramat Gan 52900,
Israel }

\author{I.\ V.\ Protopopov}
\affiliation{
 Institut f\"ur Nanotechnologie, Karlsruhe Institute of Technology,
 76021 Karlsruhe, Germany
}
\affiliation{
 L.\ D.\ Landau Institute for Theoretical Physics RAS,
 119334 Moscow, Russia
}

\author{D.\ B.\ Gutman}
\affiliation{Department of Physics, Bar Ilan University, Ramat Gan 52900,
Israel }
\affiliation{
 Institut f\"ur Nanotechnologie, Karlsruhe Institute of Technology,
 76021 Karlsruhe, Germany
}

\begin{abstract}
We  study wave  propagation in interacting Bose liquid, where
the short range part of the  interaction between atoms is of a  hard core type,  and  its long range part   
scales  with a distance as a power law.  The cases of Coulomb, dipole-dipole  and 
Van der Waals interaction are considered. We employ a hydrodynamic approach,  
based on the exact solution of Lieb-Liniger model, and  study  the  evolution of  a density pulse
instantly released from  potential trap.
We analyze  semi-classical  Euler and continuity equations and construct the corresponding Riemann invariants. We supplement our analysis with  numerical calculations and  discuss experimental applications  
for  ultacold atom experiments. 
\end{abstract}

\maketitle

\section{Introduction}
\label{s1}
Study of  interacting one dimensional systems experienced its renaissance in the last decade.
A significant progress was made both on the  experimental and the theoretical sides.
The experimental developments were boosted by new  techniques in measurements and fabrication,
in the solid state\cite{Amir}  and  ultra-cold atoms\cite{Bloch_review,Ozeri,Ketterle} systems.
The later employ optical traps,  allowing   an unprecedented level of control over the system.
Tuning the system in and out of Feshbach  resonance determines  the strength of  inter-atomic interaction, while  changing optical  traps controls the confining potential.
This allows to study the systems both in and out of  thermal equilibrium, 
where many interesting  physical effects were observed.
Quite remarkably,  cold atom experiments  can be performed both with fermionic
or bosonic constituencies.
This  offers an  opportunity to further explore the influence of a quantum statistics on  correlated one dimensional transport, and in particular on  the Bose-Fermi duality.
The later was originally  established  for equilibrium  Luttinger liquids (LL), but was
considerably extended in the recent years. This generalization emerged as 
 one dimensional quantum liquids were studied  beyond the LL paradigm\cite{Imambekov11}.
The key point  of the LL theory is Dirac spectrum of single particle fermionic excitations.  
This amount to linear hydrodynamic description, and in particular dispersionless  wave propagation.  
A generic  single-particle spectrum, on other hand,  leads to  a non-linear  quantum hydrodynamic.
In other words, finite curvature of fermionic  spectrum  generates  an interaction terms between the bosonic modes. This to be contrasted with original electrostatic interaction that induces the curvature of the 
in the bosonic branches of excitations.
From the bosonic point of view the picture is absolutely dual, as one can (re)-fermionize the problem.
In this case the interaction of the bosonic modes induces the curvature of the fermionic excitations, 
while the curvature of the bosonic spectrum generate electron-electron interaction\cite{PGOM}.

The emergent Fermi-Bose duality allows for a powerful techniques, that enable 
to study the kinetic of interacting Fermi or Bose  system.
Moreover, in the presence of a finite short range interaction, the statistics of the elementary
constituencies plays only a minor role, and the system is in a universal 
Fermi-Luttinger liquid regime\cite{khodas07}.
A particular illustrative example, is a problem of a pulse evolution, 
induced by an instant release of a fluid from a confining potential.
%Desptite being irrelevant perturbations from the renormalization group point of view,
%these bosonic vertices  lead to significant physical effects.
%In particular, they control a thermolization and inelastic scattering rate, 
%as well as time dependent evolution of a density pulse,  induced  by a local quench in  fermionic fluid \cite{Protopopov2012}. 
Within the semiclassical  approximation the density pulse 
splits into left- and right-moving parts, short time  after it was created. These parts 
separate and move in the opposite directions. If  the spectrum of single particle excitations
were exactly linear,  the  shape of the left and right moving pulses would remain constant. 
However, any finite curvature of the single particle spectrum induces a  non-linearity, that  leads to a  deformation of the pulse and unavoidably  its''overturn''.   
While this problem was thoroughly studied for fermionics fluids\cite{Wiegmann,PGOM,Eldad,Protopopov2012},  
much remains unknown for  the bosonic ones.  We address this question in the current work.
We show that if the density pulse is sufficiently weak, the system is in a universal Fermi-Luttinger regime.
Its evolutions is identical to the one studied in \cite{PGOM,Protopopov2012}, up to a redefinition 
of the bare  parameters. For pulses, that are larger than the energy scale set by a short range interaction, 
the bosonic character of elementary particles is important and the system is in a new regime.
Below we study the evolution of the system in both limits, in the  presence of a generic long range interaction.

%%%%%%%%%%%%%%%%%%%%%%%%%%%%%%%%%%%%%%%%%%%%%%%%%%%%%%%%%%%%%%%%%%%%%%%%%%%
\begin{figure}[b]
 \includegraphics[width=200pt]{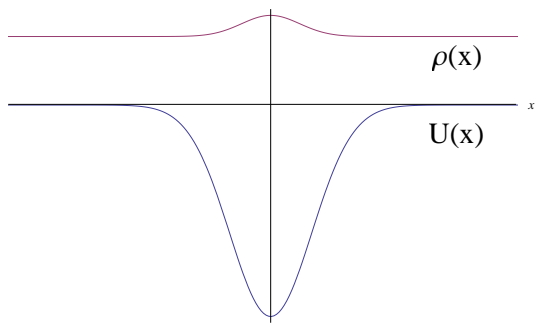}
\caption{\small Setup. Initial density disturbance  is
created by the application of a potential $U(x)$ which is then switched off at
$t=0$.  }
\label{Ur}
\end{figure}
%%%%%%%%%%%%%%%%%%%%%%%%%%%%%%%%%%%%%%%%%%%%%%%%%%%%%%%%%%%%%%%%%%%%%%%%%%%%%%

The structure of this paper is as follows.
We begin our discussion by  formulating the microscopic model, building on a Lieb-Liniger model.
In Sec. \ref{s2} we develop its hydrodynamic description  and construct 
the corresponding Riemann invariant\ref{Riemann}.
In Sec.\ref{long_range}  we analyze the behavior of the pulse evolution in the presence of generic long range interaction.
We supplement the analytic arguments with numeric calculation.
In Section \ref{conclusion} we summarize our result and discuss further extensions.

\section{Hydrodynamic approach to  Lieb-Liniger model}
\label{s2}

%The study of 1D boson interactions has started for almost century ago, but there are still plenty of challenging problems to be solved. Some of them had been solved exactly, some have only approximate 1D models and solutions. For years, researchers mainly focused their attention on 1D electron systems. Much less attention has been given to the 1D bosons. In fact, the one can easily show that the free fermion model exactly describes bosons, interacting through the two-body potential $V_{int}(x)$.

We start our analysys by introducing  a microscopic model, described by the following Hamiltonian    
\begin{equation} 
\hat{H}=\sum_{i=1}^{N}\left(\frac{\hat{p}_i^2}{2m}+V_{\rm ext}(x_i)\right)+\sum_{i<j=1}^{N}V_{\rm int}(x_i-x_j)\,.
\label{eq:Initial Hamiltonian}
\end{equation}
Here $V_{\rm ext}$ is a confining potential and $V_{\rm int}$ is a two-particle interaction, 
that consist of a short and a long range parts
\begin{equation}
V_{\rm int}(x)=g'\delta(x)+V_{\rm lr}(x)\,.
\end{equation}
Let us first consider the case where only short range interaction is present
 (we will relax this assumption later on).
In the  limit the system is described    Lieb-Liniger model\cite{lieb-liniger,giamarchi}
\begin{equation} 
\hat{H}=\sum_{i=1}^{N}\left(-\frac{\hbar^2}{2m}\frac{\partial^2}{\partial{x}_{i}^2}+V_{\rm ext}(x_i)\right)+ g'\sum_{i<j=1}^{N}\delta({x_i}-{x_j}) \,.
\label{eq:LL Hamiltonian}
\end{equation}
%There are several models describing 1D bosons, while the simpliest and a non-trivial is Lieb-Liniger model, that assumes zero external potential $ V_{ext}(x)$ and the two-body Dirac-delta interaction $ V_{int}(x)=g\delta(x)$. 
It is convenient to parametrize the interaction strength in this model by
\begin{equation}
c =\frac{mg'}{ h^2} ,
\end{equation}
that  has a dimension of an inverse length. Thus, $ c = 0$ limit  corresponds to free bosons, while $c \to \infty$ is the hard-core or Tonks-Girardeau limit. In this case the 
density correlation function of any order  coincide with the one  of   non-interacting fermions.
Even for a finite  values of interaction strength $c$ the Lieb-Liniger  problem is exactly solvable.
The problem of pulse evolution for this model was recently solved in Ref.\cite{Andrei,Garry}.
Moreover,  this limit was analyzed in details within the hydrodynamic approach\cite{Hoefer,Damski,Abanov}.  
Since we are interested in the situation where both   long and short range part of interaction are present, 
we   consider the known case of  Lieb-Liniger model  first.

To develop a unified approach it is convenient  to pass into the collective degrees of freedom, 
"bosonizing"  the bosonic Hamiltonian (\ref{eq:Initial Hamiltonian}).  
By doing so, we restrict ourself  to  the case of the smooth potential trap. 
%aiming at hydrodynamic description of the problem.  
In terms of hydrodynamic variables the Hamiltonian is given by
\begin{equation} 
\hat{H}=\int dx \left[\frac{m}{2}\hat{\rho}{\hat{v}^2}+E(\hat{\rho})+\frac{\hat{\rho}_x^2}{8m\hat{\rho}}\right]\,.
\label{eq:Hydrodynamic Hamiltonian}
\end{equation}
Here $\hat{\rho}$ is a density field, $\hat{v}$ is a velocity field and 
$E(\hat{\rho})$ is  an energy per unit  volume.

Now on we limit ourself to quasi-classical accuracy, neglecting the difference between an operator and 
its expectation value.  The equations of motions of Hamiltonian \ref{eq:Hydrodynamic Hamiltonian} became
the continuity and Euler equations for an ideal Bose fluid.
The replacement of the true field theory by classical equation of motion, is certainly an approximation.
It is justified under a number of conditions:
(a) the shape of the pulse is sufficiently smooth; (b) the pulse contains a large number of particles in it;
(c) we consider the system at time scales shorter than the inelastic scattering time. 
(d) the long-range part of the interaction is sufficiently strong. 
In this work we will focus on the case when all the assumptions above are satisfied.

Assuming that the  hydrodynamic fields are slow function of space and time
we  use an  exact solution of Lieb-Liniger model at equilibrium, for a local  energy  of the fluid 
per unit volume \cite{giamarchi}
\begin{equation} 
E(\rho)=\frac{\hbar^2}{2m}\rho^3e(\gamma)\,.
\label{eq:Energy}
\end{equation}
Here  
\begin{equation} 
\gamma=\frac{c}{\rho} 
\label{eq:gamma}
\end{equation}
is a dimensionless interaction strength,  
%\begin{equation} 
%c=\frac{mg}{\hbar^2}\, 
%\label{eq:c}
%\end{equation}
%is the sound velocity.
and  the function  $e(\gamma)$  is known from  the Bethe ansatz solution.  
It encodes the information about the interaction on a short scale  and determines the  thermodynamic properties of  the liquid.
In particular,  using \ref{eq:Energy} one finds the pressure in the fluid
\begin{equation}
\label{pressure}
P=\frac{\rho^3 e(\gamma)}{m}-\frac{c\rho^2}{2m}\frac{\partial e(\gamma)}{\partial \gamma}\,.
\end{equation}
For strong values of interaction strength  ($ \gamma \to \infty$),  the function  $e(\gamma)$  approaches  a constant value ($e(\gamma)=\frac{\pi^2}{3}$),  reproducing the free fermions limit.
For weak values of interaction  ($ \gamma\leq 1$)  
\begin{equation} 
e(\gamma)\simeq\gamma[1-\frac{4}{3}{\pi}{\sqrt{\gamma}}+\dots] 
\label{eq:Bogoliubov approximation}
\end{equation}
in agreement with  the Bogoliubov approximation\cite{lieb-liniger}.
The Hamiltonian \ref{eq:Hydrodynamic Hamiltonian} together with Eq.\ref{eq:Energy}   lead to the
continuity 
%\begin{equation} 
%\hat{\rho}_{t}=i[\hat{H},\hat{\rho}] 
%\label{eq: initial commutator}
%\end{equation}
\begin{equation} 
\partial_{t}\rho+\partial_{x}(\rho v)=0\,
\label{eq:continuity equation}
\end{equation}
and  Euler equation 
\begin{equation} 
\partial_{t}v+v\partial_{x}v+\partial_{x}w=0\,.
\label{eq:Enthalpy in Euler equation}
\end{equation}
Here   
\begin{equation}
w(\rho)=w_0(\rho_0)-\frac{1}{4}\partial_x^2\ln \rho-\frac{1}{8}(\partial_x\ln\rho)^2+\dots
\end{equation} 
is the gradient expansion of enthalpy per unit mass.
Its leading part  is given by
\begin{eqnarray}
w_0(\rho)=\frac{3\hbar^2}{2m^2}\rho^2e(\gamma)-\frac{\hbar^2c\rho}{2m^2}\frac{\partial e(\gamma)}{\partial \gamma}\,.
\end{eqnarray}
In the limiting of the strong interaction ($\gamma \gg 1$) one finds
\begin{equation} 
w_0\simeq\frac{\pi^2\hbar^2\rho^2}{2m^2}\left(1- \frac{4}{3\gamma} + \dots\right)
\label{eq:strong interection enthalpy}
\end{equation}
In the limit of  the weak interaction ($\gamma\ll1$)
\begin{equation}
w_0\simeq \frac{\hbar^2c\rho}{m^2}\left(1-\pi\sqrt{\gamma} +\dots \right)
\label{eq:weak interection enthalpy}
\end{equation}

Depending on the average  density, the strength of the pulse and interaction the system can be in one of the following regimes.
If the amplitude  of the pulse is small  
$  (\Delta \rho)^2/2m \ll \rho g' \leq \rho^2/ 2m$ the system  is in the  universal Fermi-Luttinger liquid regime.
In this case, up-to replacement of the parameters  (i.e. sound velocity, effective mass and interaction constants) 
the evolution of the pulse is similar to the one found in a fermionic model\cite{Protopopov2012}. 
If the amplitude of the pulse is big, i.e.   case $ (\Delta \rho)^2/2m \geq \rho g'$ the system is a qualitatively  new regime,
where bosonic nature of its constituencies is distinct. In the current work we explore the transition between these two regimes.

We now proceed with the analysis of the pulse evolution,  first  limiting ourself to the finite range interaction.
In this case, as it was shown by Riemann,  classical  hydrodynamic equation  unavoidably  leads to the formation of a shock wave. The  position  of the shock and the behavior of the solution in its vicinity depend on the "regularization". 
In  the absence of long-range interaction the  regularization arise due to the terms with high order spatial derivatives in Hamiltonian \ref{eq:Hydrodynamic Hamiltonian}. Away from the shock this derivatives are small, and can be neglected.
Close the shock, this term start to play a role, stabilizing the solution, and leading to oscillation in the density and velocity fields. This oscillation takes place of the  short scale,  indicating that the assumptions of a smooth solution do not  hold and hydrodynamic approach is no   longer valid.  
Therefore, it makes it no sense to study the evolution of  pulse in the hydrodynamic approach unless long range interaction is present.  Before we include this part of the interaction into consideration, 
let us stay with a Lieb-Liniger model a bit longer and  calculate  the time when the shock wave is  formed and  its position.  In order to do it, we construct Riemann invariants.

\subsection{Riemann invariants}
\label{Riemann}
As it was pointed out by Riemann,  a flow of an ideal  one dimensional liquid  preserves  two constants of motion.  The are called  Riemann invariants, and   are  constructed  as  follows \cite{stone,Landau}
\begin{equation} 
J_\pm={v}\pm\int\frac{dP}{m\rho{}s}\,,
\label{eq:Riemann invariants}
\end{equation}
%\begin{equation} 
%\int\frac{1}{\rho{}s}dP,
%\end{equation}
where $s=s(\rho)$ is the speed of sound, and the functions $P$ and $\rho$ are related to each other via Eq.(\ref{pressure}).

In terms of the Riemann invariants the equation of motion can be writen as\cite{stone,Landau}
\begin{equation}
\bigg[\frac{\partial}{\partial t}+(v\pm s)\frac{\partial}{\partial x}\bigg]J_\pm=0\,.
\label{eq:Equation of motion in terms of Riemann invariant}
\end{equation}
%The Euler and continuity equations in terms of the above integral become so called Riemann equations that are given by\cite{Dima}:

%\begin{equation} 
%{\partial{}}_t\left(v+\int\frac{1}{\rho{}s}dP\right)+\left(v+s\right){\partial{}}_x\left(v+\int\frac{1}{\rho{}s}dP\right)=0
%\end{equation}

%\begin{equation} 
%{\partial{}}_t\left(v-\int\frac{1}{\rho{}s}dP\right)+\left(v-s\right){\partial{}}_x\left(v-\int\frac{1}{\rho{}s}dP\right)=0
%\end{equation}

This implies  that $J_{\pm}$
are constants along the characteristic curves
\begin{equation} 
\frac{dx(t)}{dt}=v(x(t),t){\pm}s(x(t),t).
\label{eq:Characteristic curves}
\end{equation}
Here $s$  is sound  velocity,  determined by  
\begin{equation}
s=\frac{1}{\sqrt{m}}\sqrt{\left(\frac{\partial P}{\partial \rho}\right)_S}\,,
\label{eq:General sound velocity}
\end{equation}
that can be easily evaluated using Eq.(\ref{pressure}).

\begin{figure}[htp!]
%\centering
\includegraphics[width=210pt]{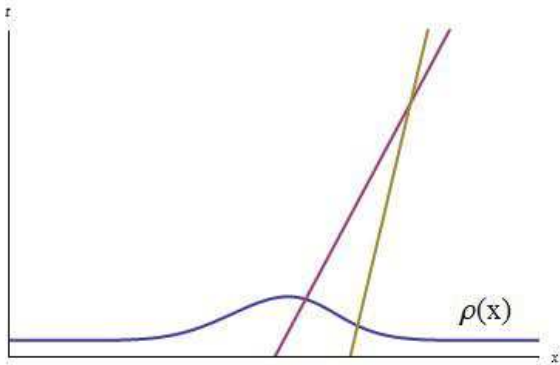}
\caption{schematic illustration for the characteristics in a unilaterally  propagating pulse}
\label{fig4}
\end{figure}

Eq.(\ref{eq:Characteristic curves}) is  highly non-trivial, since it described  the motion of the particles with  coordinate $x(t)$, moving  in an  unknown velocity and density fields.
However,  it greatly simplifies for the problem of a pulse evolution, once 
its  left and right moving are separated in space. 
As it was rigorously shown by Riemann \cite{Landau},  in this case the  characteristics are  straight lines, see Fig. \ref{fig4}.

%For the gas  of bosons with short range interaction the equation of state
%\begin{equation}
%P V^\alpha=const
%\label{eq:Equation of state}
%\end{equation}
Now we construct the Riemann invariant for the Lieb-Liniger model.
For the   strong interaction  limit ($\gamma \gg 1$)    the pressure is  
\begin{equation} 
P(\rho)=\frac{\pi^2\hbar^2}{3m}{\rho^3}\,,
\label{eq:Strong interaction P(rho)}
\end{equation}
that correspond to a  gas with a polytropic index  $n=3$.
The  sound  velocity in this limit is 
%\begin{equation}
%s=\sqrt{\left(\frac{\partial P}{\partial \rho}\right)_S}
%\label{eq:General sound velocity}
%\end{equation}
\begin{equation}
\label{sound_strong} 
s(\rho)=\frac{\pi\hbar}{m}\rho\,,
\end{equation}
and the Riemann invariants are
\begin{equation} 
J_\pm=
%{v}\pm\int\frac{1}{m\rho s}dP
{v}\pm\frac{\pi\hbar}{m}(\rho-\rho_0)\,.
\label{eq:Riemann invariants strong case}
\end{equation}

For the weak interaction
% $\int\frac{1}{\rho{}s}dP$ is proportional to $\sqrt{\rho}$ with  
the pressure is given by
\begin{equation} 
P(\rho)=\frac{c\hbar^2}{2m}{\rho^2},
\label{eq:Weak interaction P(rho)}
\end{equation}
corresponding to a gas with the polytropic index  $n=2$. 
%when $c$ here is the interaction strengh.
The speed of sound in this limit reads
\begin{equation} 
s(\rho)= \sqrt{{\frac{c\hbar^2\rho}{m^2}}},
\label{sound_weak}
\end{equation}
and the corresponding Riemann invariants are
\begin{equation} 
J_\pm= v\pm \frac{2\hbar\sqrt{c}}{m}(\sqrt{\rho}-\sqrt{\rho_0}).
\label{eq:Riemann invariants weak case}
\end{equation}
Using this analysis one can easily find the time $t_c$ when the shock wave is formed.
%%%%%%%%%%%%%%%%%%%%%%%%%%%%%%%%%%%%%%%%%%%%%

%\begin{figure}[t]
%\label{shock}
%\centering
%\includegraphics[width=210pt]{wave_before_shock}
%\caption{Density profile at $t$ slightly smaller than the "shock'' time $t_c$. Only right-moving part is shown.The front edge of the pulse has become steep but the overturn has not yet occurred.}
%\end{figure}

%%%%%%%%%%%%%%%%%%%%%%%%%%%%%%%%%%%%%%%%%%%%%

It   occurs when two the characteristic curves intersect
\begin{equation} 
t_c =\Delta x /\Delta s.
\label{eq:General equation for the shock time}
\end{equation}
Here  $ \Delta x$ is the width of the pulse,  $\Delta \rho$ is the difference in the density between the pulse 
and the unperturbed region, $\Delta s$ is the difference of the sound velocity  at the maximum of the pulse and away from it.

%%%%%%%%%%%%%%%%%%%%%%%%%%%%%%%%%%%%%%

%%%%%%%%%%%%%%%%%%%%%%%%%%%%%%%%%%%%%%

Using  Eqs. (\ref{sound_strong}, \ref{sound_weak}) one finds

\begin{equation} 
t_c =\frac{m}{\pi \hbar} \frac{\Delta{x}}{{\pi}{\Delta \rho}},
\label{eq:equation for the shock time strong}
\end{equation}
for $\gamma \gg 1$ and
\begin{equation} 
t_c =\frac{\Delta x}{\Delta \rho}\frac{m}{\hbar}\sqrt{\frac{\rho}{c}}
\label{eq: equation for the shock time weak}
\end{equation}
for  $\gamma \ll 1$.  One sees that the time of the shock formation $t_c$   is parametrically longer for for   weakly  interacting  Bosons compared with strongly  interacting ones.

\section{Long range interaction}
\label{long_range}
We consider three types of long range interaction, that are most relevant 
for the experiments: Coulomb, dipole and Van der Waals, accounted by 
\begin{equation}
H_{\rm int}=\frac{1}{2}\int dx dx' V(x-x')\rho(x) \rho(x')\,
\end{equation}
with a corresponding choice of an interaction potential $V$.
In this case  the  hydrodynamic equations (\ref{eq:continuity equation},\ref{eq:Enthalpy in Euler equation}) hold, but 
with en entalpy term is modified 
\begin{equation}
w(x) \rightarrow w(x)+\int dx'  V(x-x') \rho(x')\,.
\end{equation}
Because  all  three  types of interaction  depend on a distance as power, 
we  first discuss  a generic power law interaction $V_{\rm }(x) \sim 1/x^\alpha$.
A competition between interaction induced dispersion and non-linear terms induces a density modulations, 
with a characteristic scale  $\Delta x$.
The oscillations  are pronounced in the region, where both terms are of the same order
\begin{equation}
v^2 \sim l_0^{\alpha-2} \Delta \rho \Delta x^{1-\alpha}\,.
\end{equation}
 The relation between the density and  the velocity part in the  right  moving wave can be found  from the condition
$J_-=0$,  and similarly to the left moving pulse.
For the case $\gamma \gg 1$  this estimate  yields  $ v \sim \pi\rho/m$, that  leads to the 
\begin{equation}
\Delta x \sim \frac{l_0}{(\Delta \rho l_0)^\beta}, \,\,\,\beta =\frac{1}{(\alpha-1)}\,
\end{equation}
in agreement with Ref. \cite{Protopopov2012}.
In the case $\gamma \ll 1$ similar estimate  yields $v \sim \Delta \rho \sqrt{c\rho_0}/\rho_0$.
The corresponding scale of oscillation is given by
\begin{equation}
\Delta x \sim l_0 \left(\frac{m\rho_0}{c\Delta \rho l_0}\right)^\beta\,.
\end{equation}
In a time domain, one expects an appearance of a new pick, that emerges from a bump, approximately 
each $t_c$.  This estimate holds as long a size of the bump is of the order of the original one. 
At times when  the size and the shape are significantly deformed, this estimate is no longer valid. 

We now focus on typical cases of inter-atomic interaction, starting with  
Coulomb interaction ($\alpha=1$). 
We model it by a screened Coulomb  potential  
\begin{eqnarray}
V(x)=\frac{1}{ml_0}\left(\frac{1}{\sqrt{x^2+a^2}}-\frac{1}{\sqrt{x^2+d^2}}\right)\,.
\end{eqnarray}
Here $a$ is the  transversal  size of the trap, and $d$ is the distance to the screening gate.
In this case, the long range part of the interaction in the Fourier space is given by
\begin{eqnarray}&&
V_{\rm lr}(q)
%g'+\frac{2}{ml_0}\left(K_0(qa)-K_0(qd)\right)=\nonumber \\&& 
=\frac{2}{ml_0}\left(K_0(qa)-K_0(qd)-\ln(d/a)\right).
\end{eqnarray}
Here $K_0$ is modified Bessel function,
and the effective amplitude of the short range potential is 
$g=g'+2/ml_0\ln d/a$.
The renormalization of the bare value of the short range interaction $g'$ occur  due to a singularity 
of Coulomb potential at small distances. 
%We model   two particle interaction, e.g.   Eq.(\ref{eq:Initial Hamiltonian}), by a sum
%of a short range and a long range parts
%\begin{equation}
%V(r)=g'\delta(r)+ V_{\rm lr} (r) \,.
%\end{equation}

%%%%%%%%%%%%%%%%%%%%%%%%%%%%%%%%%%%%%%%%%%
\begin{figure*}[ht!]
\includegraphics[width=210pt]{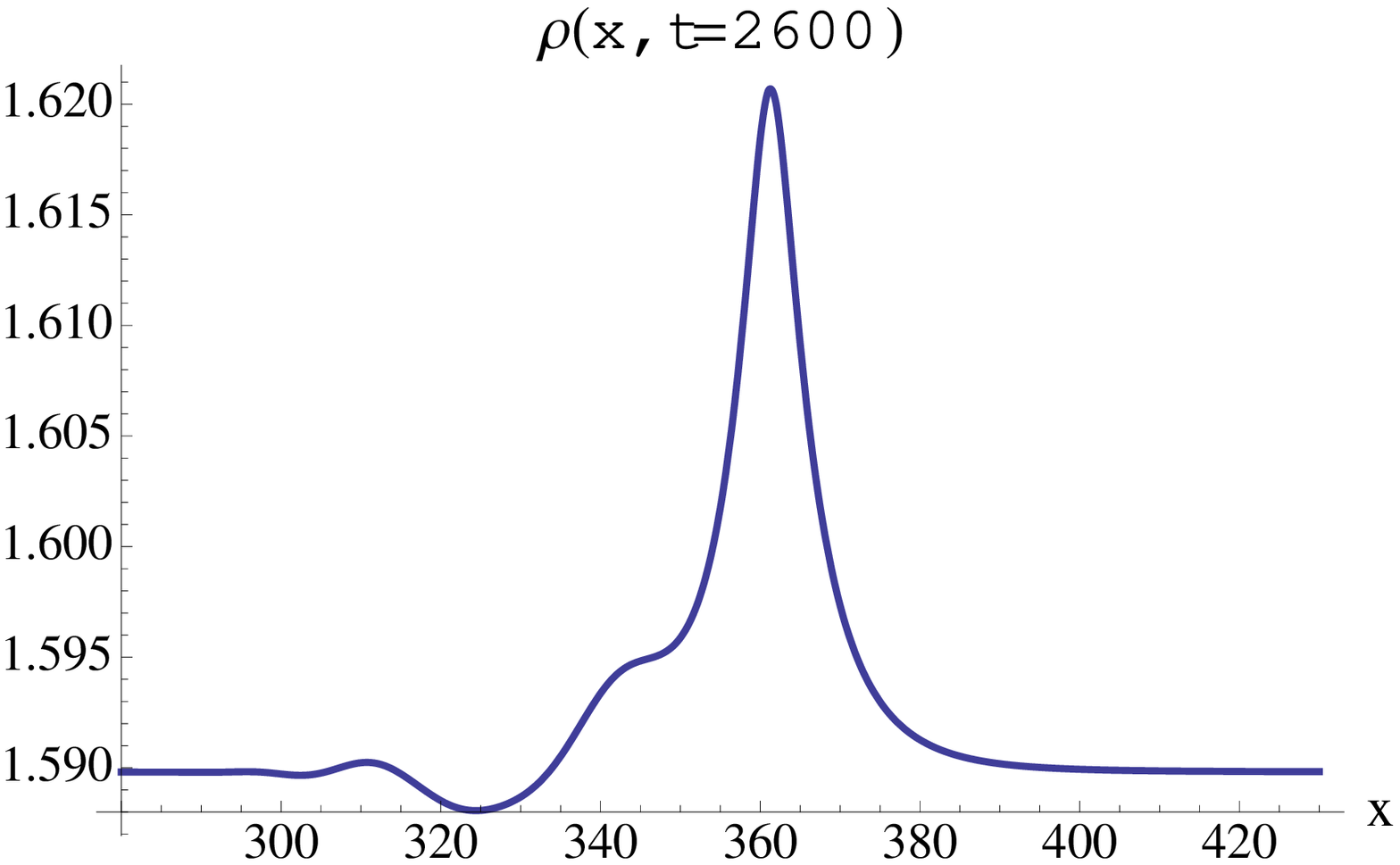}
\includegraphics[width=210pt]{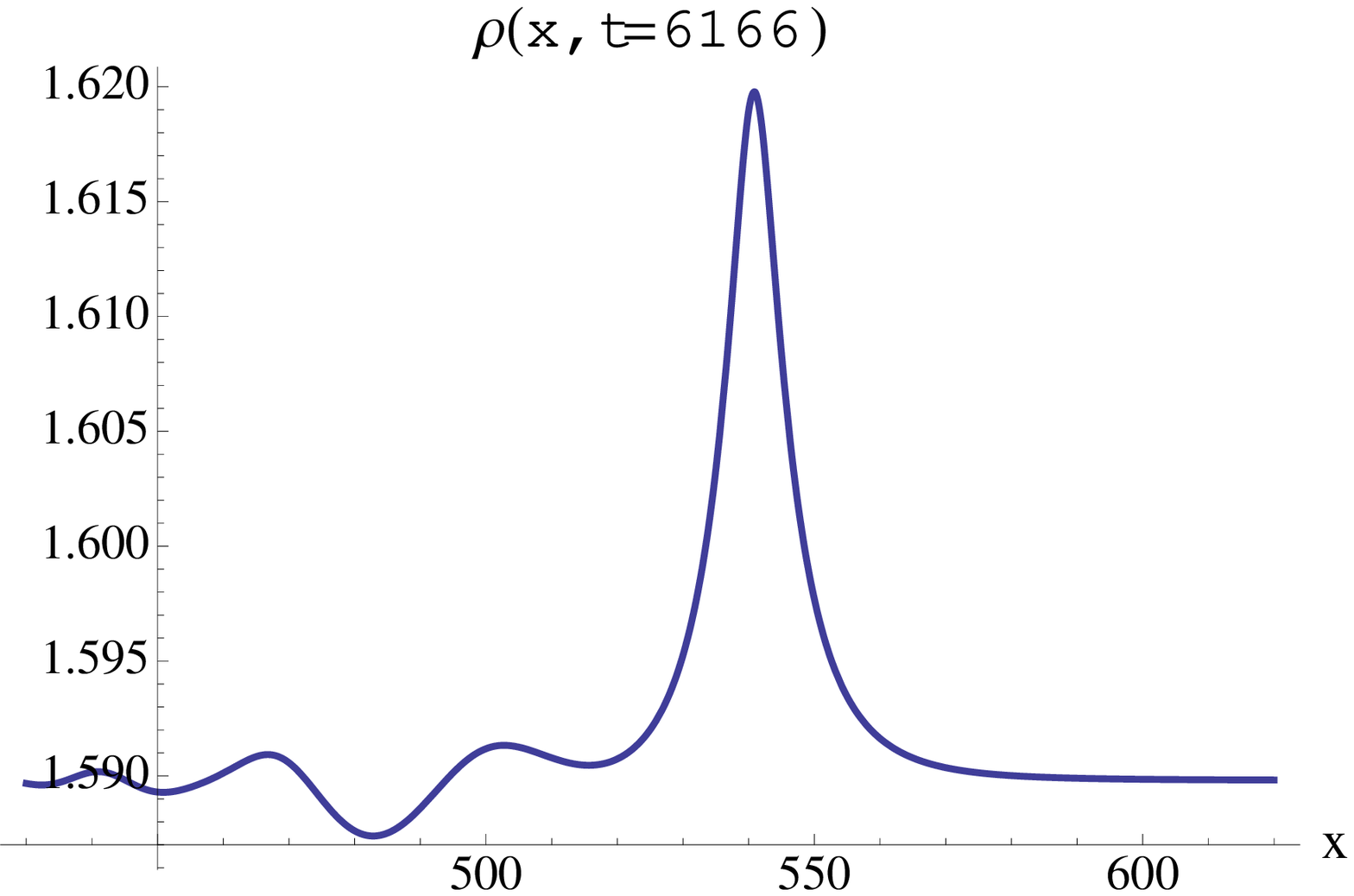}
\includegraphics[width=210pt]{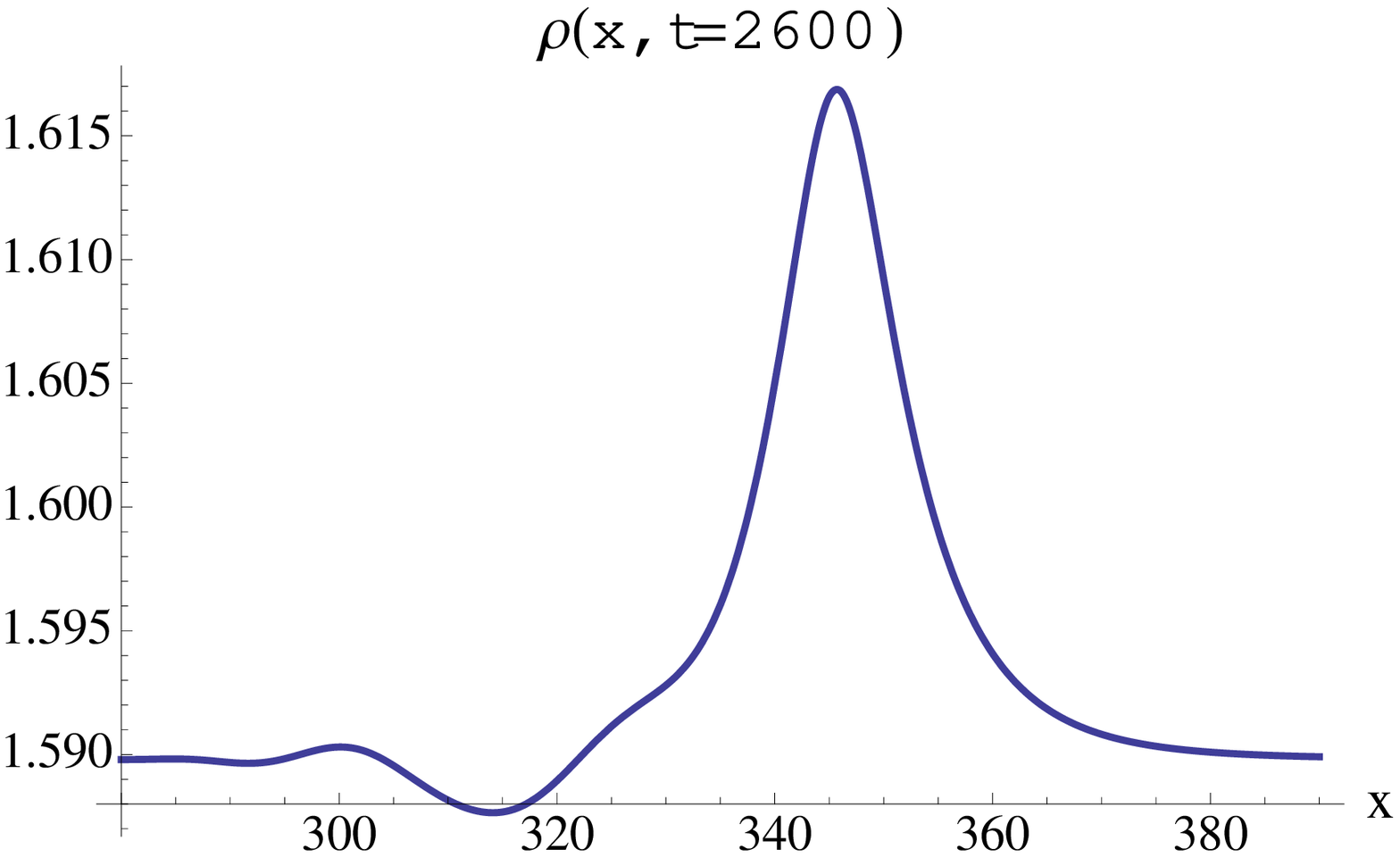}
\includegraphics[width=210pt]{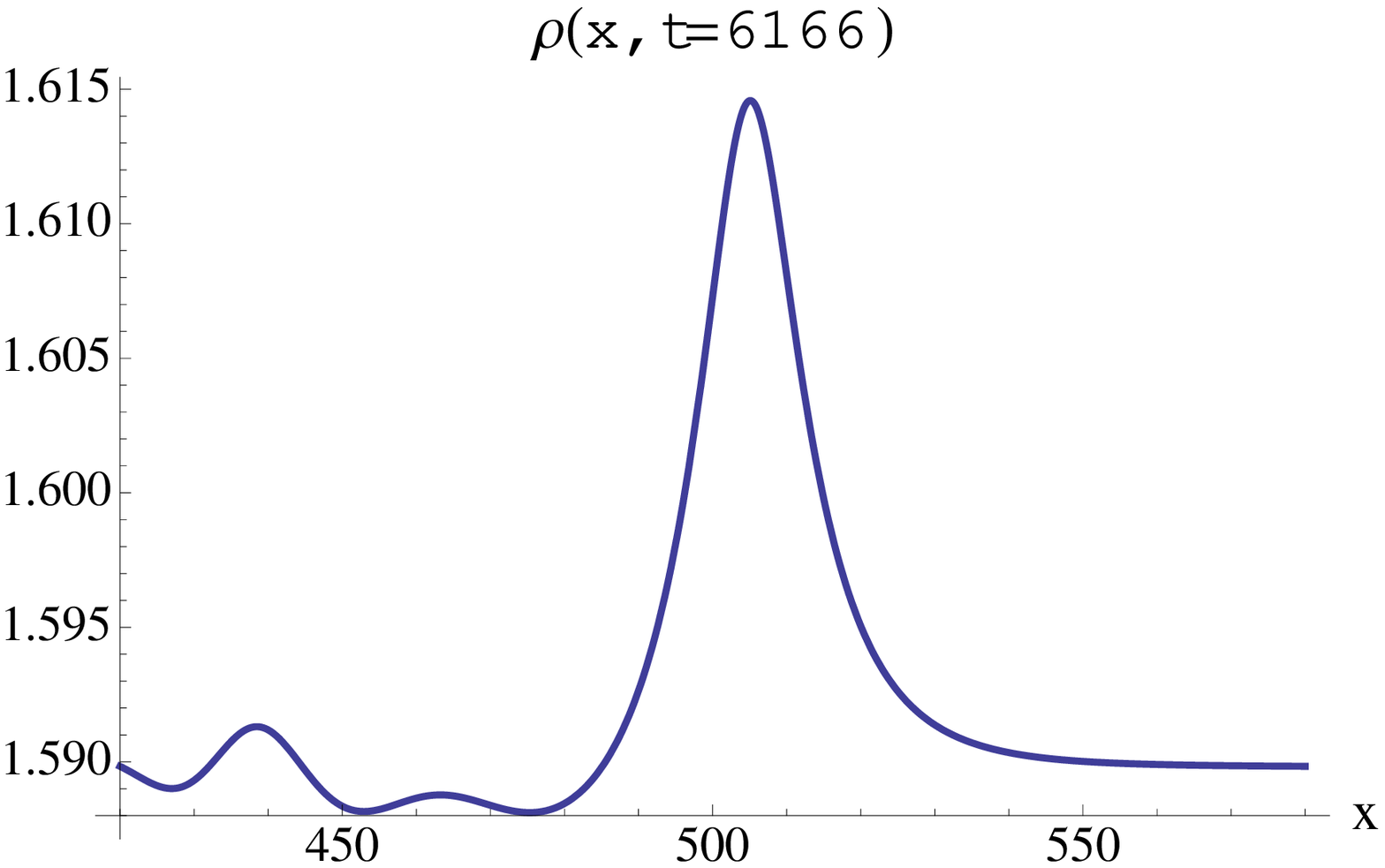}
\caption{\small  
Density profile after the shock for the case of Coulomb interaction, $\gamma \gg 1$ (top) and 
$\gamma \ll 1$ (bottom). }
\label{Coulomb_d}
\end{figure*}

\begin{figure*}[ht!]
\includegraphics[width=210pt]{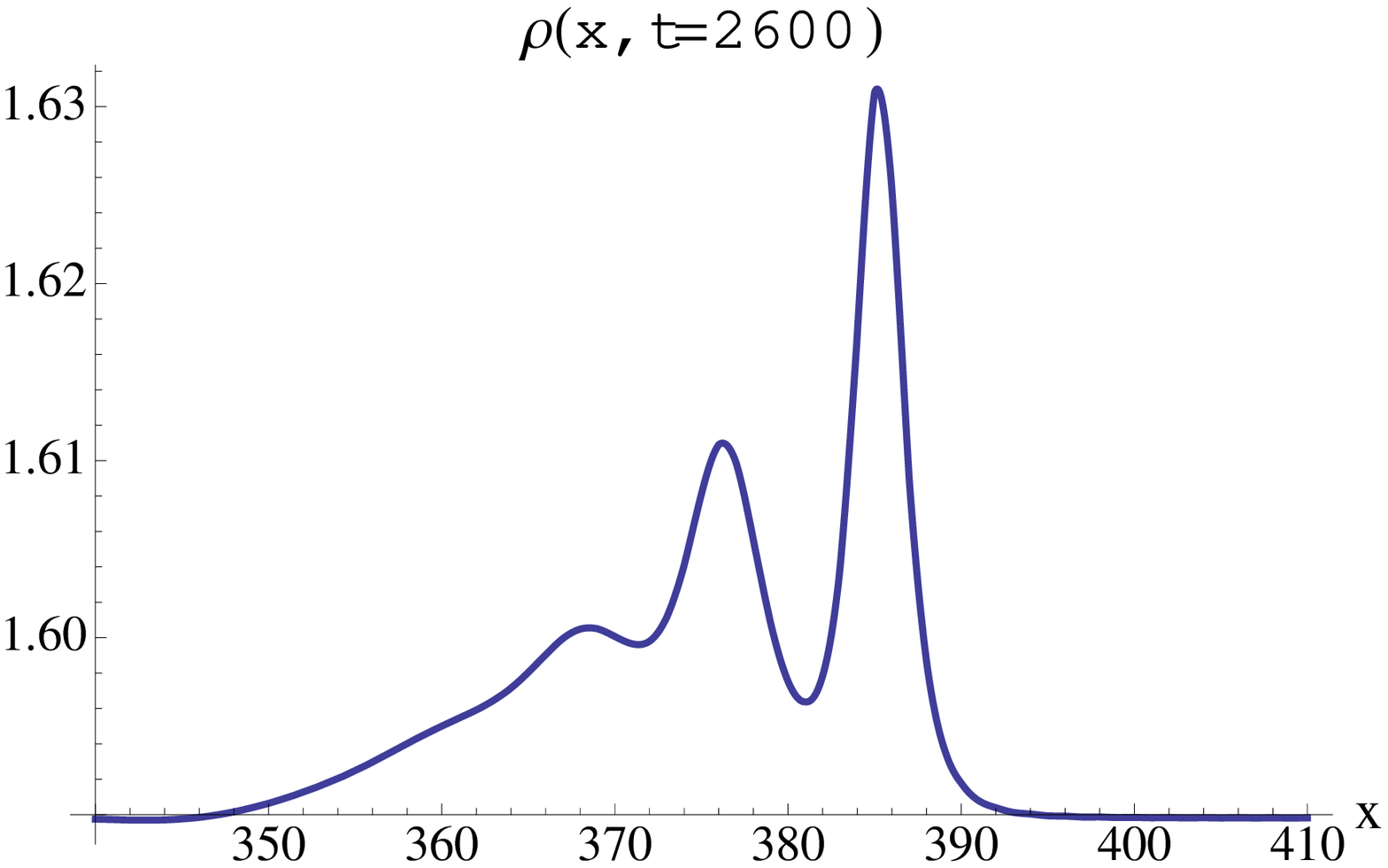}
\includegraphics[width=210pt]{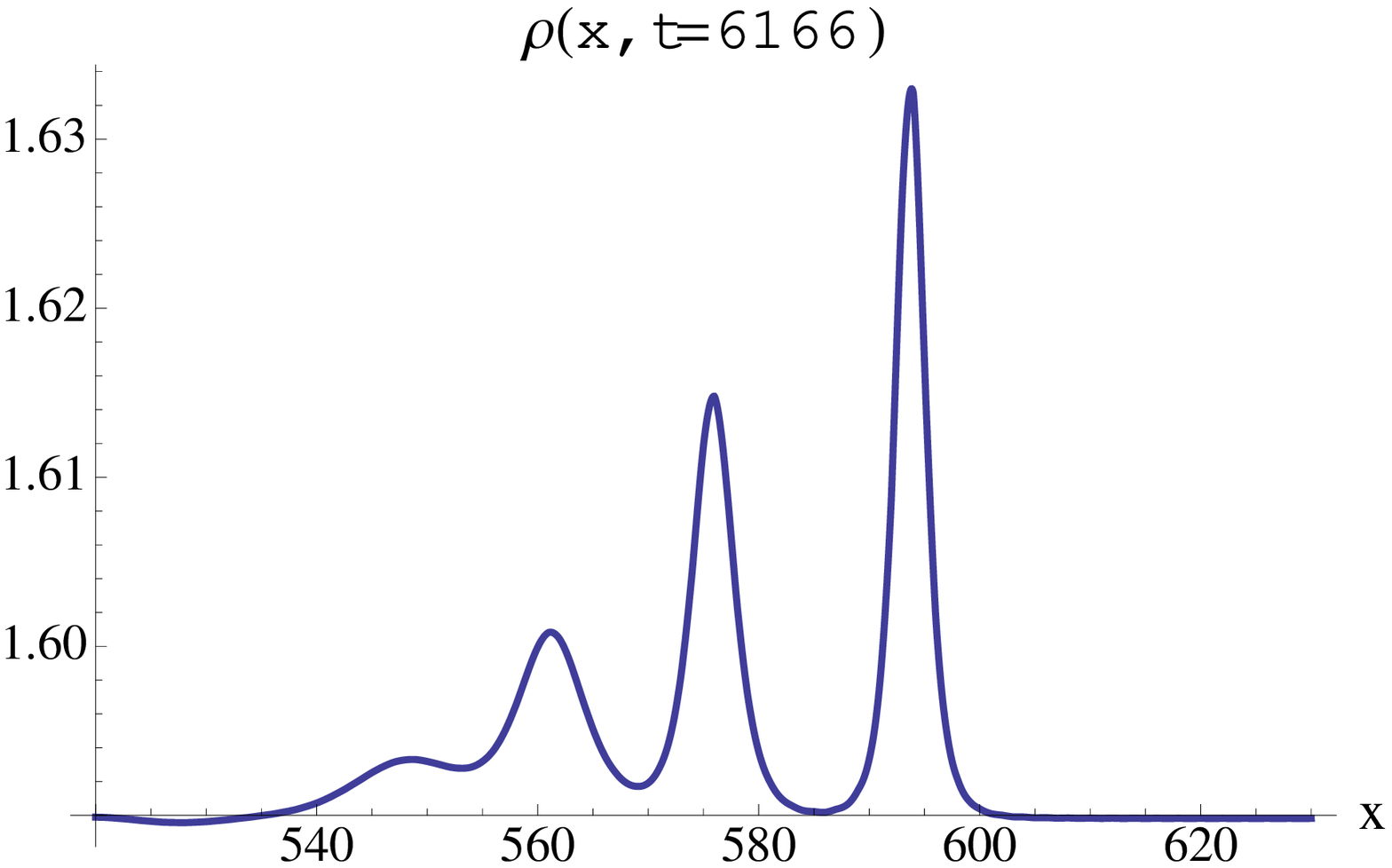}
\includegraphics[width=210pt]{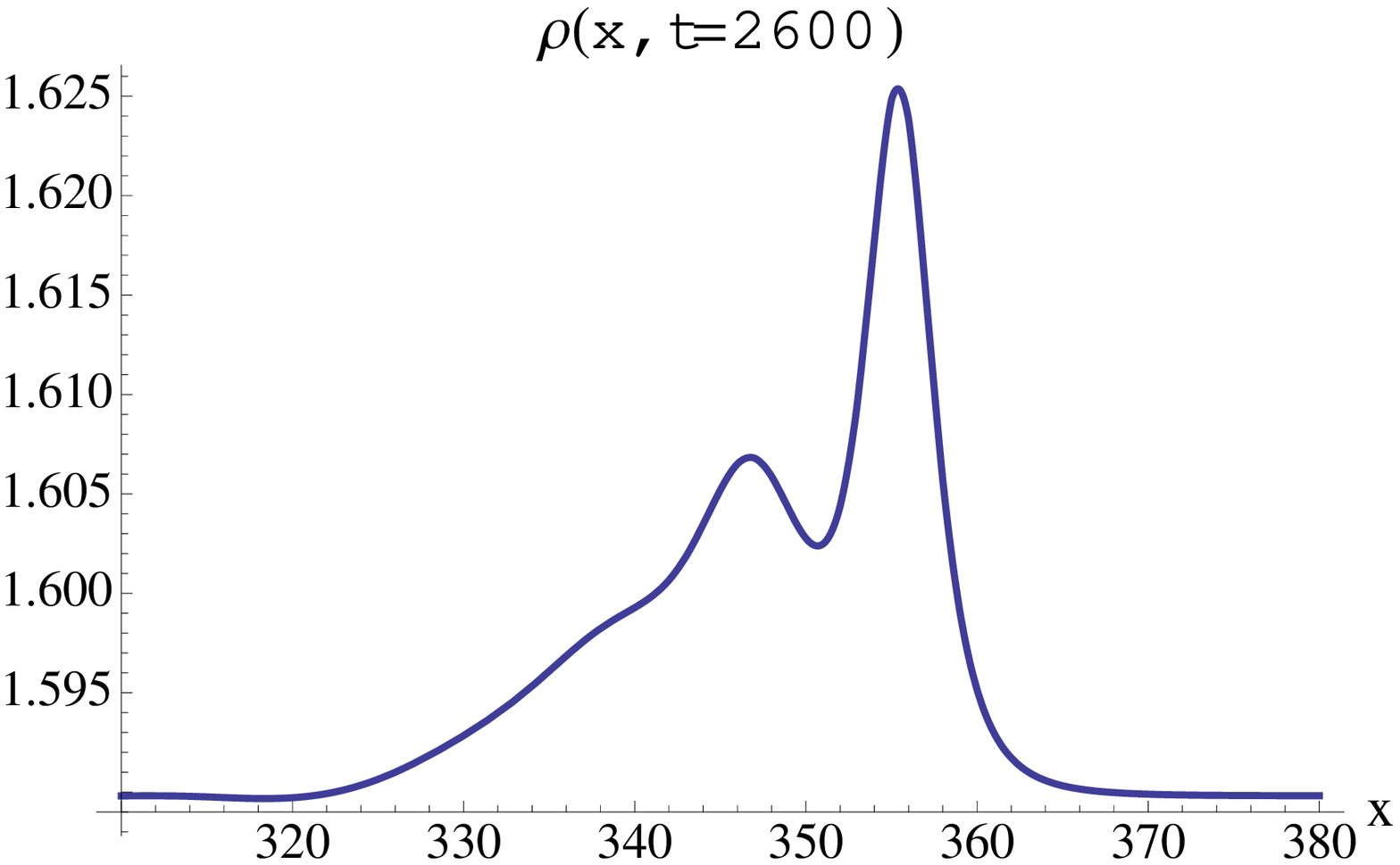}
\includegraphics[width=210pt]{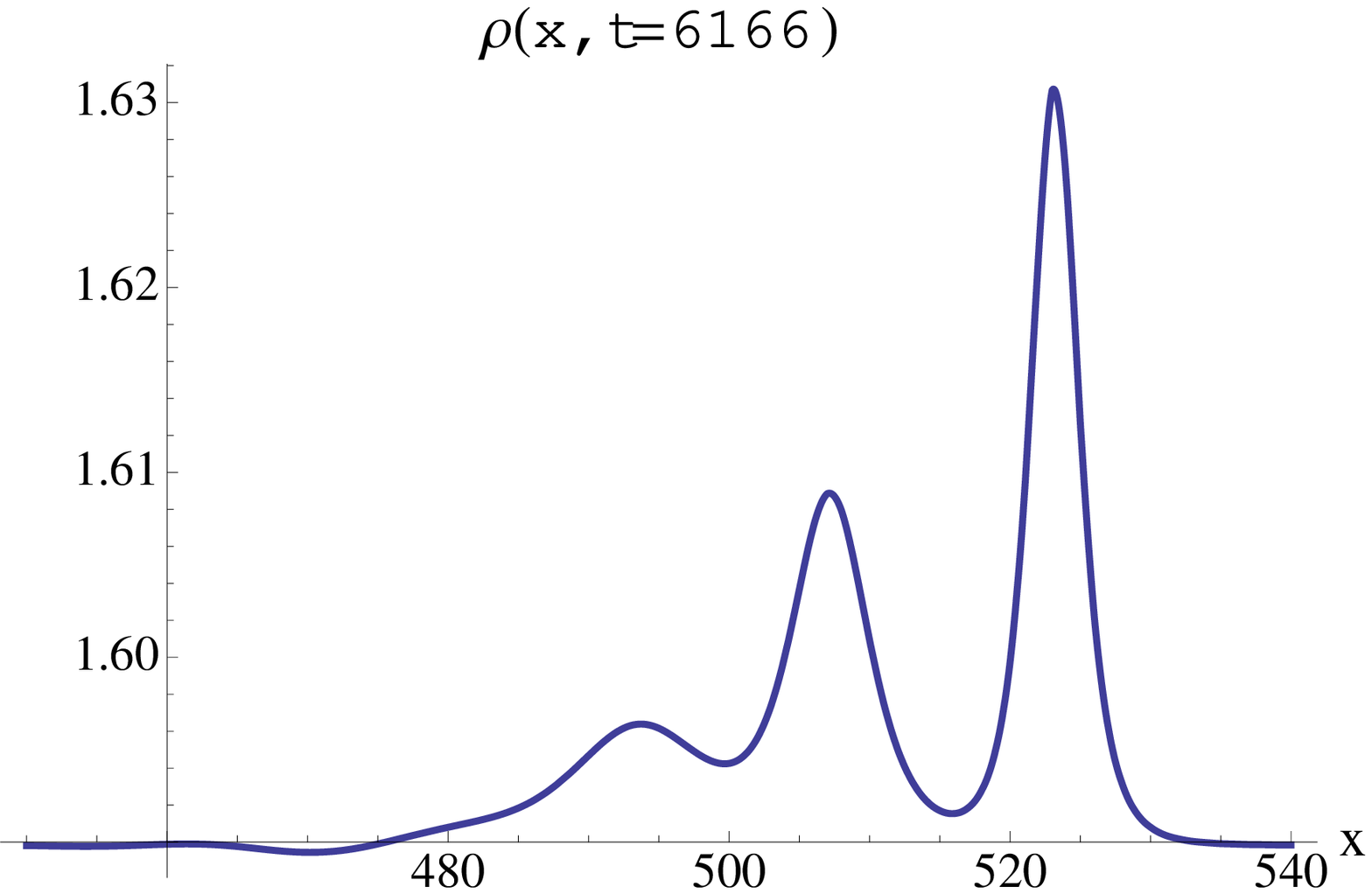}
\caption{\small  
Density profile after the shock for the case of dipole  interaction, $\gamma \gg 1$ (top) and $\gamma \ll 1$ (bottom)}
\label{dipole}
\end{figure*}

\begin{figure*}[ht!]
\includegraphics[width=210pt]{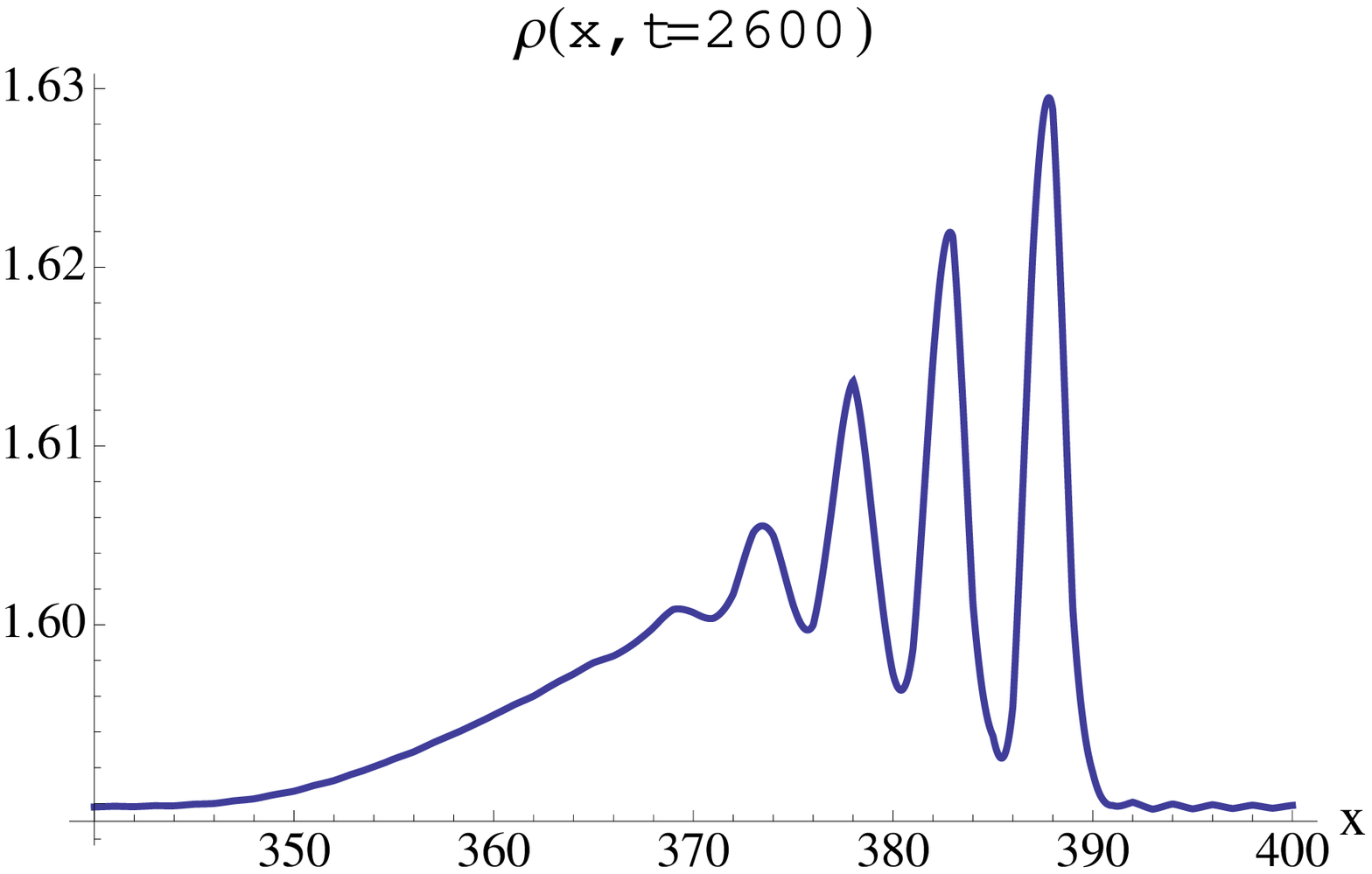}
\includegraphics[width=210pt]{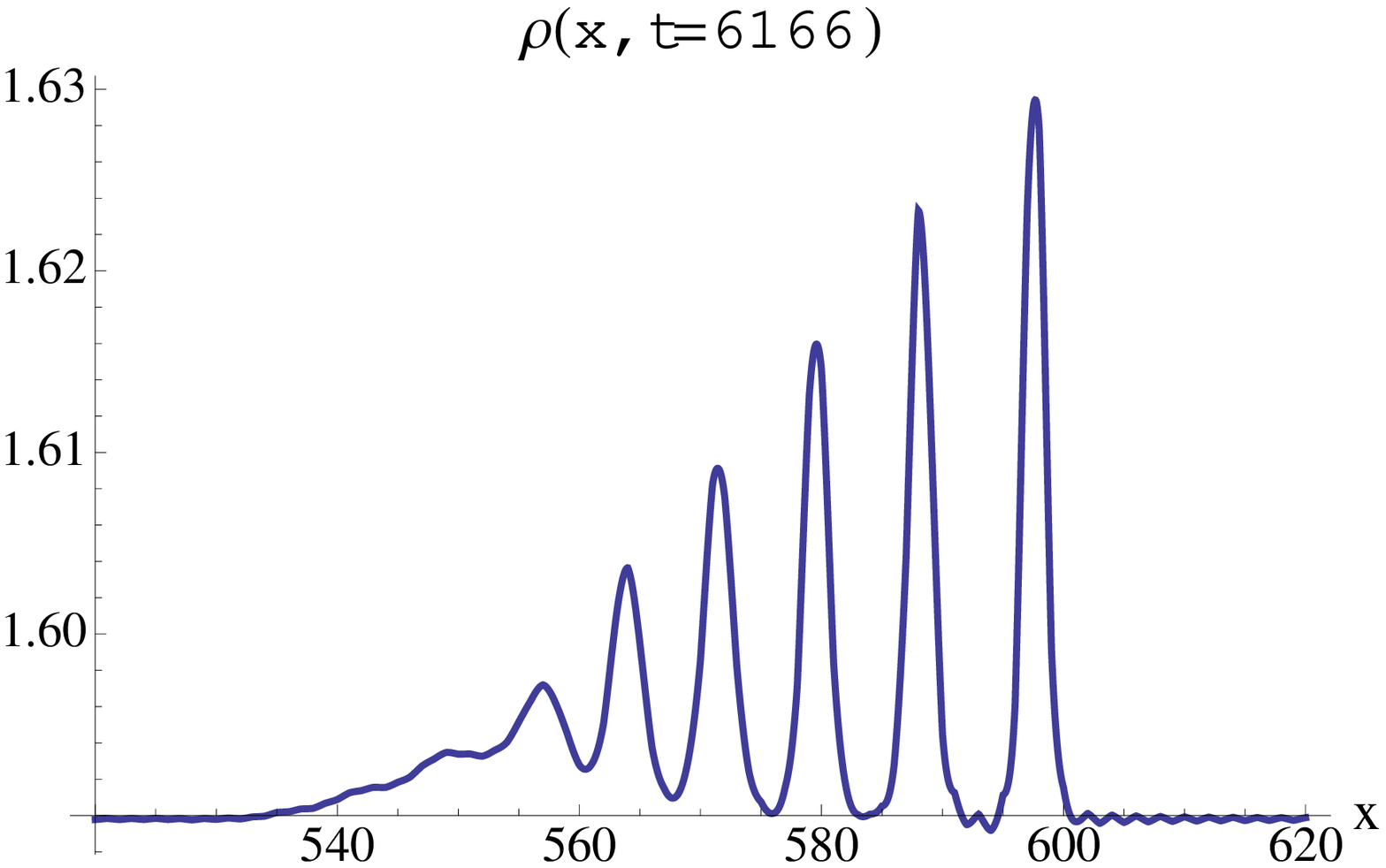}
\includegraphics[width=210pt]{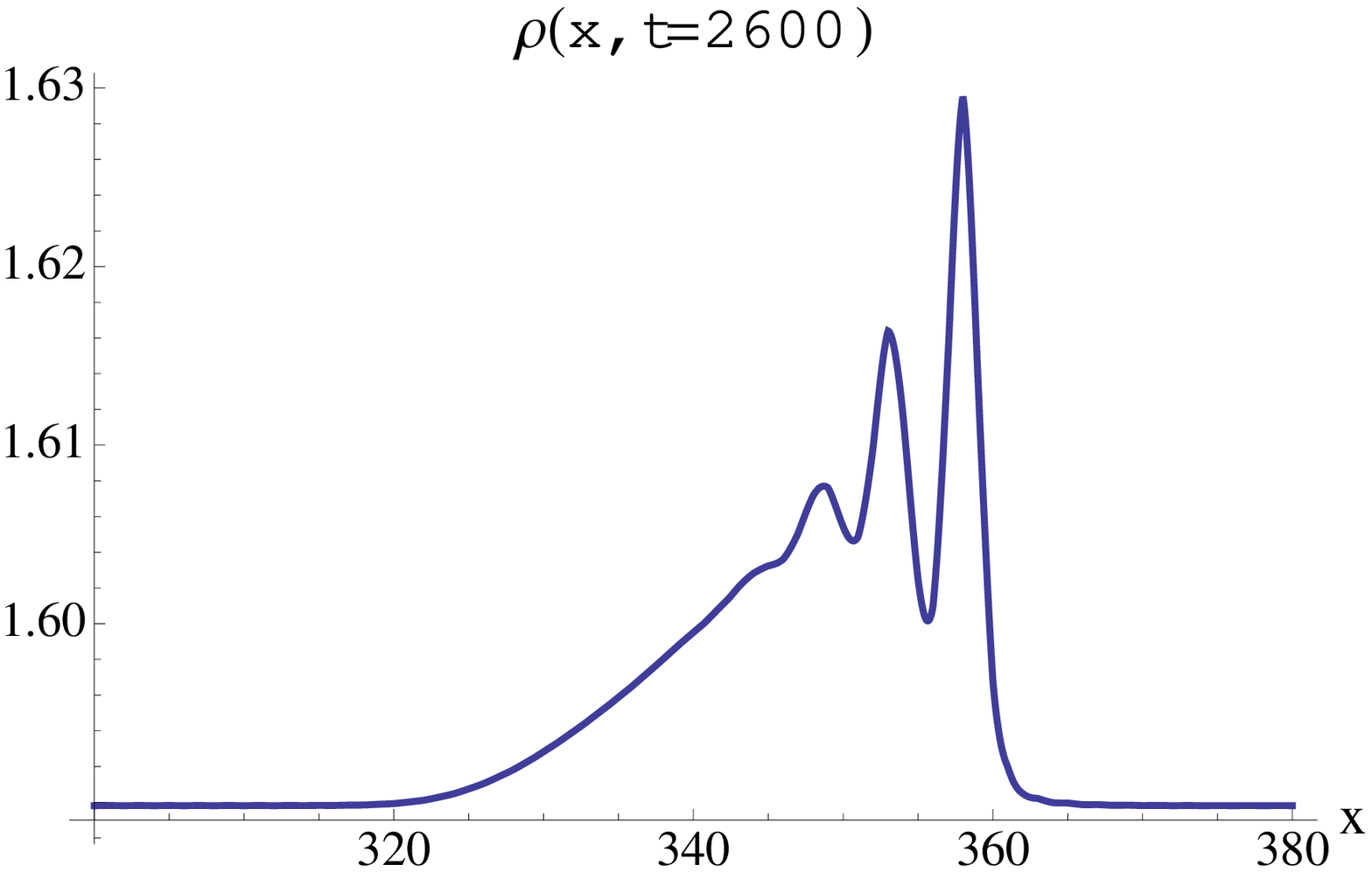}
\includegraphics[width=210pt]{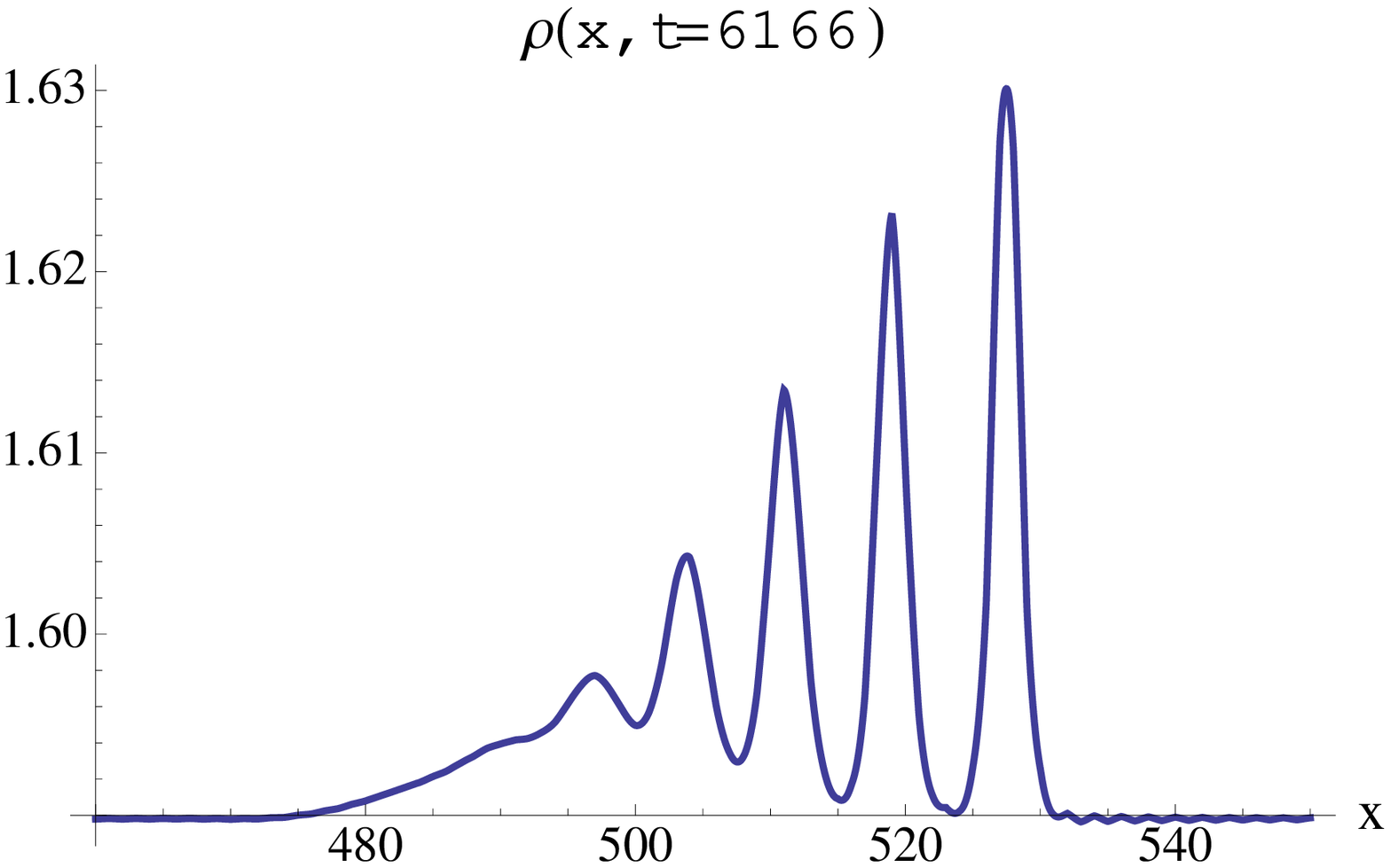}
\caption{\small  
Density profile after the shock for the case of Van der Waals   interaction, $\gamma \gg 1$ (top) and $\gamma \ll 1$ (bottom)}
\label{Van_der_Waals}
\end{figure*}

%%%%%%%%%%%%%%%%%%%%%%%%%%%%%%%%%%%%%%%%%%%%%

%\begin{figure*}[h]
%\centering
%\includegraphics[width=210pt]{shock_wave_regularized}
%\caption{Wave evolution  with Coulomb regularization }
%\end{figure*}

%%%%%%%%%%%%%%%%%%%%%%%%%%%%%%%%%%%%%%%%%%%%%

%%%%%%%%%%%%%%%%%%%%%%%%%%%%%%%%%%%%%%%%%%

%\begin{figure*}[!ht]
%\centering
%\includegraphics[width=210pt]{Coulomb_strong_int}
%\includegraphics[width=210pt]{Coulomb_weak_int.eps}
%\caption{Wave evolution of Bose gas with Coulomb repuslion: $\gamma \gg 1$ on the left and  $\gamma \ll 1$ on the right}
%\label{fig2}
%\end{figure*}

%%%%%%%%%%%%%%%%%%%%%%%%%%%%%%%%%%%%%%%%%%

%%%%%%%%%%%%%%%%%%%%%%%%%%%%%%%%%%%%%%%%%%

%\begin{figure*}[h]
%\centering
%\includegraphics[width=210pt]{Coulomb_weak_int.eps}
%\caption{Wave evolution of Bose gas with weak ($\gamma \ll 1$) Lieb-Liniger  interaction }
%\label{fig3}
%\end{figure*}

%%%%%%%%%%%%%%%%%%%%%%%%%%%%%%%%%%%%%%%%%%

%%%%%%%%%%%%%%%%%%%%%%%%%%%%%%%%%%%%%%%%%%

%\begin{figure*}
%\includegraphics[width=210pt]{Calogero_weak_int.eps}
%\includegraphics[width=210pt]{Calogero_strong_int.eps}
%\caption{\small  
%Evolution of the bosonic density for Calogero interaction.  Both of the figures show the density at $t \approx 5t_c$.
%{\it Left:} The density profile for weak short range interaction ($\gamma\approx 1$).
%{\it Right:} The density profile for weak short range interaction ($ \gamma \to \infty$) .   
% }
%\label{Calogero}
%\end{figure*}

%%%%%%%%%%%%%%%%%%%%%%%%%%%%%%%%%%%%%%%%%%%

The hydrodynamic equation can be now simulated on the computer, leading 
to the density evolution shown in the Fig.\ref{Coulomb_d} .
%To perform the calculations we impose the periodic boundary conditions,  and
%employ the  fast Fourier transform algorithm for the calculation of $\partial_{t}\rho$. 
%Combined with the standard fourth-order Runge-Kutta time stepper this provides us with the fast and accurate algorithm 
%for the numerical solution of Eq.({\ref{eq:Enthalpy in Euler equation}).
Comparing  the evolution of density profile in two limiting cases we note, 
that  the pulse for $\gamma \gg 1$, shown in  
Fig. \ref{Coulomb_d},  has propagated a longer  distance that the one with $\gamma \ll 1$.
This corresponds  to a bigger sound velocity in the former limit as expected.
However, the difference in the distance, that the pulse has propagated,   and its  shape 
is rather mild.  This has to do with a strong renormalization of the bare value of a short range interaction ($g'$), by a Coulomb potential. The difference between the  renormolized values of short rage interaction
 $g$ is not big. As a result, for Coulomb interaction the system remains effectively in the Fermi-Luttinger limit, even though it was not a case for original Lieb-Liniger model.
The main pick in Fig.\ref{Coulomb_d} is followed by an oscillatory tail.

%The results for the  Calogero potential for the studied cases are shown on Fig.\ref{Calogero}  for the large and small values of $\gamma$ respectively.
If atoms are neutral, the direct Coulomb interaction we have just discussed is absent.
In this case, the leading part of a long range interaction is determined by dipole-diploe interaction ($\alpha=3$).
Known examples are ultracold  chronium\cite{dipole2, dipole3} and dysposium\cite{dipole1} atoms.
Tuned to a state with a   non-zero dipole moment  these atoms interact via
\begin{equation}
V(x)=\frac{C_3}{(|x|+a)^3}\,.
\end{equation}
In a Fourier space it  corresponds to
\begin{equation}
V_{\rm lr}(q)=C_3q^2\ln\left(qae^{\gamma_E} \right),
\end{equation}
and $g=g'+\frac{C_3}{a^2}$ for the renormalized short range interaction strength;
$\gamma_E\simeq 0,5772$ is Euler constant.
%If the long-range part of the interaction is neglected the model is equivalent to Lieb-Liniger moedl with 
%the interaction constant $g=g'+\frac{C_3}{a^2}$.
%{\bf Alex,  we need to repeat to plot the evolution with this potential as well.
% In the program notations $C_3q^2\ln\left(qae^\gamma \right)$ should stand instead of PotentialDescriptor in the program}
The evolution of the pulse in this  case is shown in Fig. \ref{dipole} for the values of weak and strong values  of a bare short range interaction.
One notice a substantial difference between the evolution of the liquid with small and large values of bare interaction. In the later case, the pulse moves with a higher  speed and  the number of  oscillations in it is larger.

%\begin{figure*}[!ht]
%\centering
%\includegraphics[width=210pt]{C3_Correlation_High_Gamma_rho.png} 
%\includegraphics[width=210pt]{C3_Correlation_Low_Gamma_rho.png} 
%\caption{density evolution for dipole-dipole  interaction, $\gamma \gg 1$ on the left and $\gamma \ll 1$ on the right}
%\label{dipole}
%\end{figure*}

%%%%%%%%%%%%%%%%%%%%%%%%%%%%%%%%%%%%%%%%%%%%%

%\begin{figure*}[!ht]
%\centering
%\includegraphics[width=210pt]{C6_Correlation_High_Gamma_rho.png} 
%\includegraphics[width=210pt]{C6_Correlation_Low_Gamma_rho.png} 
%\caption{density evolution for Van der Waals   interaction, $\gamma \gg 1$ on the left and $\gamma \ll 1$ on the right}
%\label{VanderWaals}
%\end{figure*}

%%%%%%%%%%%%%%%%%%%%%%%%%%%%%%%%%%%%%%%%%%%%%
Finally, if atoms do posses a dipole moment,  but its direction is not polarized 
the long range part of interaction is of Van der Waals type ($\alpha=6$).
This is a situation where  majority of contemporary  cold gases belong.
Because  this interaction is rather weak, it hardly influences   a scattering length $l_0$.
For this reason this interaction is usually neglected. However, it  play a major role for a dispersive regularization of the shock waves, as we show  next. 
In the real space   the  of Van der Waals  interaction  decays as
\begin{equation}
V(x)=-C_6/(x^2+a^2)^3\,.
\end{equation}
Passing to $q$ space  one finds
\begin{equation}
V_{\rm lr}(q)=+\frac{\pi C_6q^2}{8a^3}
\end{equation}
and $g=g'-\frac{3\pi C_6}{8a^5}$.
The results of pulse evolution in  this case are depicted in Fig.\ref{Van_der_Waals}.
The difference between $\gamma \ll 1$ and $\gamma \gg 1$ in this case is  stronger  pronounced here 
than for a dipole interaction, we discussed earlier.  The trend however remains the same.
The number of picks in the oscillatory tail  increases with 
increasing $\gamma$, while the amplitude of the leading pick remains practically unchanged.
One may note, that  Van der Waals  interaction induces a regularization that has  
the  same form as one induced by a quantum pressure. However,  its coefficient is  
controlled by the interaction  strength  and  does not vanish in large density limit. 
As a result, as long as a condition $\Delta \rho \ll \rho$ is satisfied Van der Walls interaction regularizes  shocks 
more efficiently  than  the quantum pressure.
Interaction  decays with distance slower than $1/x^6$  induces a regularization with 
a smaller  number of spatial derivatives. In this case it dominates  the quantum pressure term even stronger.

In addition to   density evolution,   hydrodynamic equatuions contain the information about  
the  velocity of quantum fluid in space and time. It is  quite  reasonable to dicuss it as well. 
This is what we  do it in the next section, in the context of interferometry measurements in Bose liquid.

%%%%%%%%%%%%%%%%%%%%%%%%%%%%%%%%%%%%%%%%%%%%%

%\begin{figure*}[h]
%\centering
%\includegraphics[width=210pt]{C6_Correlation_High_Gamma_rho.png} 
%\caption{density evolution for Van der Waals interaction, $\gamma \gg 1$}
%\label{fig8}
%\end{figure*}

%%%%%%%%%%%%%%%%%%%%%%%%%%%%%%%%%%%%%%%%%%%%%

%If the long-range part of the interaction is neglected the model is equivalent to Lieb-Liniger moedl with 
%the interaction constant $g=g'-\frac{3\pi C_6}{8a^5}$.
%{\bf Alex, we need  to substitute this potential instead of the Coulomb.
%In the program notations  $\frac{\pi C_6q^2}{8a^3}$
%should stand instead of PotentialDescriptor}.

\section{Interference measurements}
\label{interference}

\begin{figure*}
\includegraphics[width=210pt]{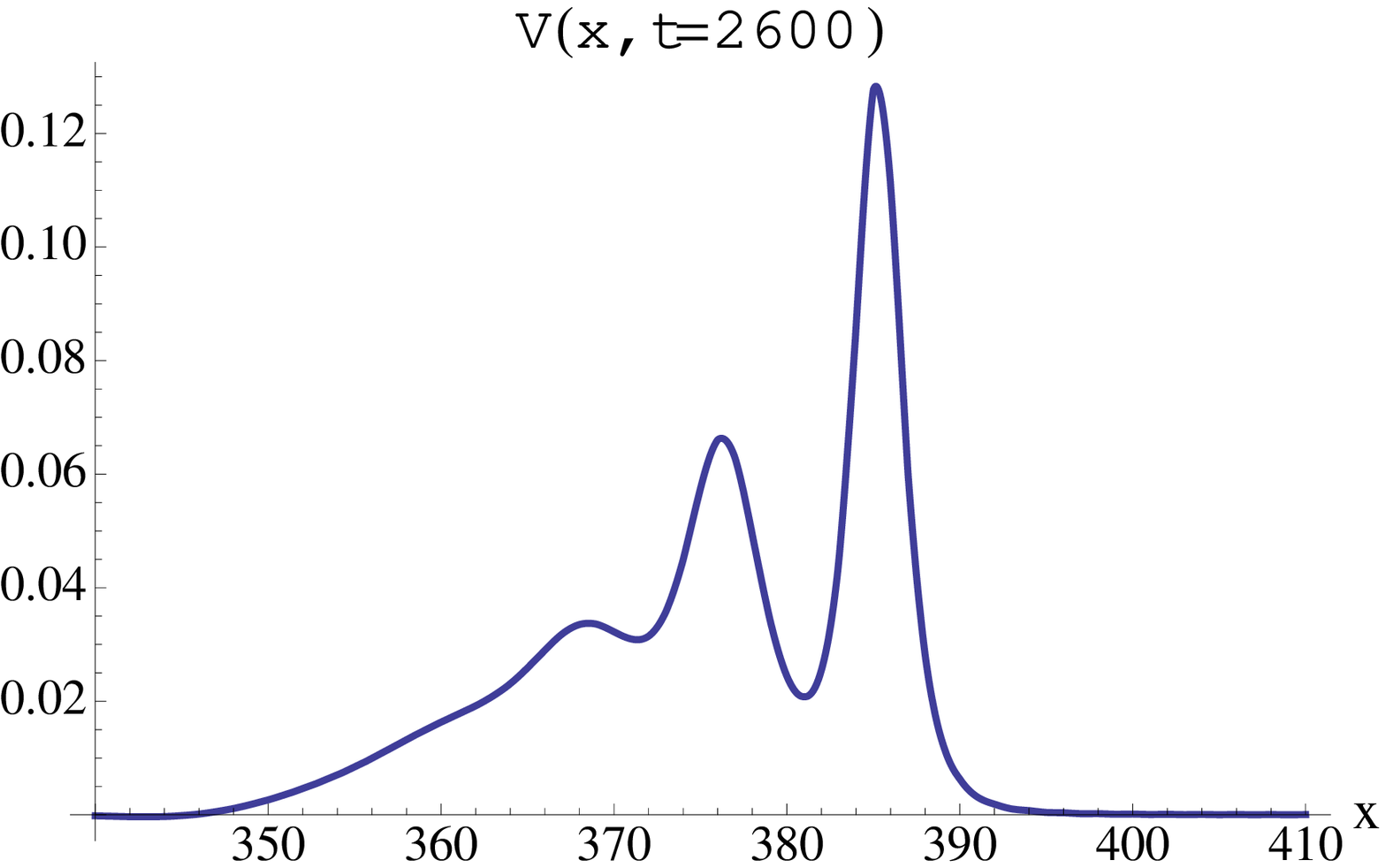}
\includegraphics[width=210pt]{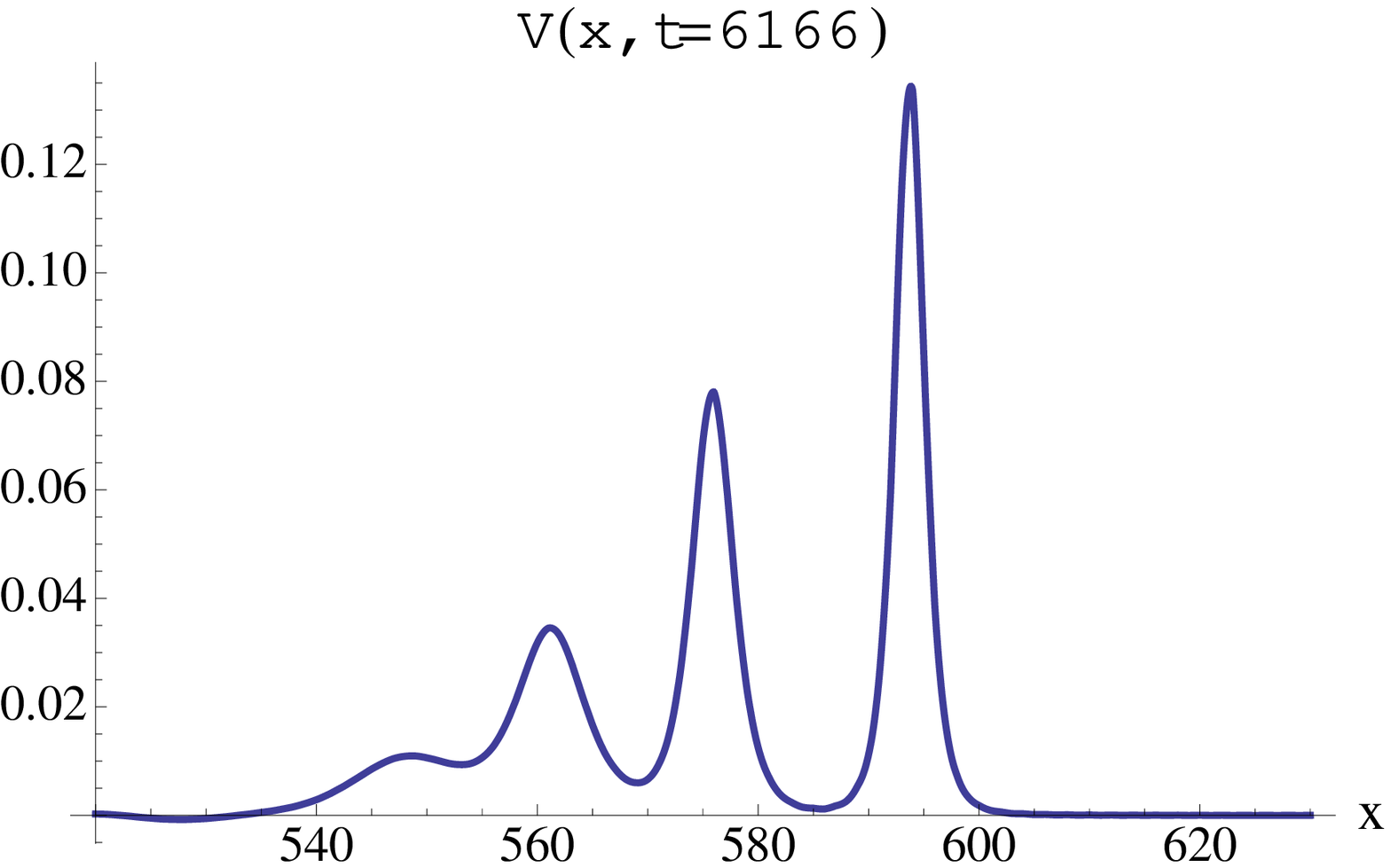}
\includegraphics[width=210pt]{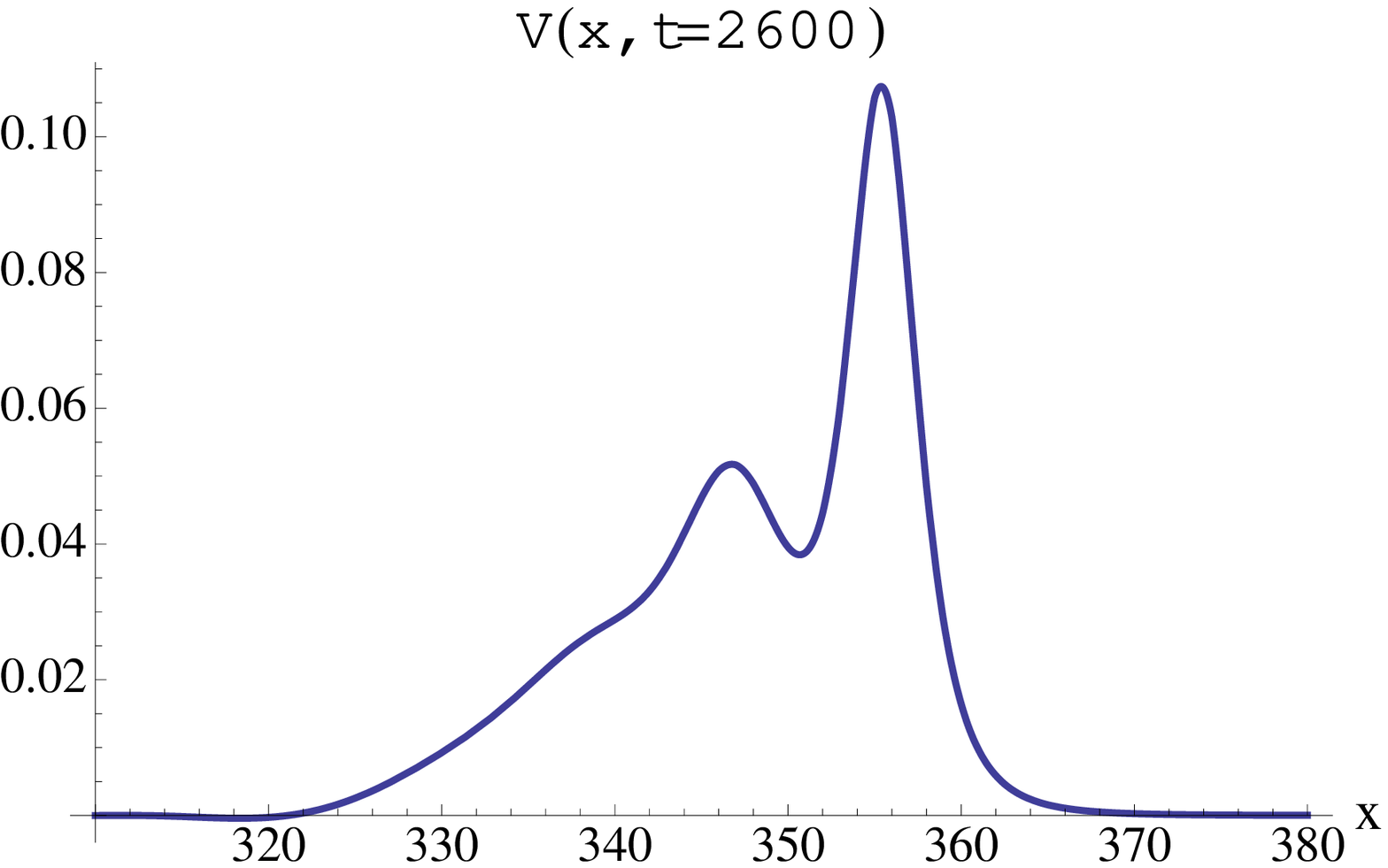}
\includegraphics[width=210pt]{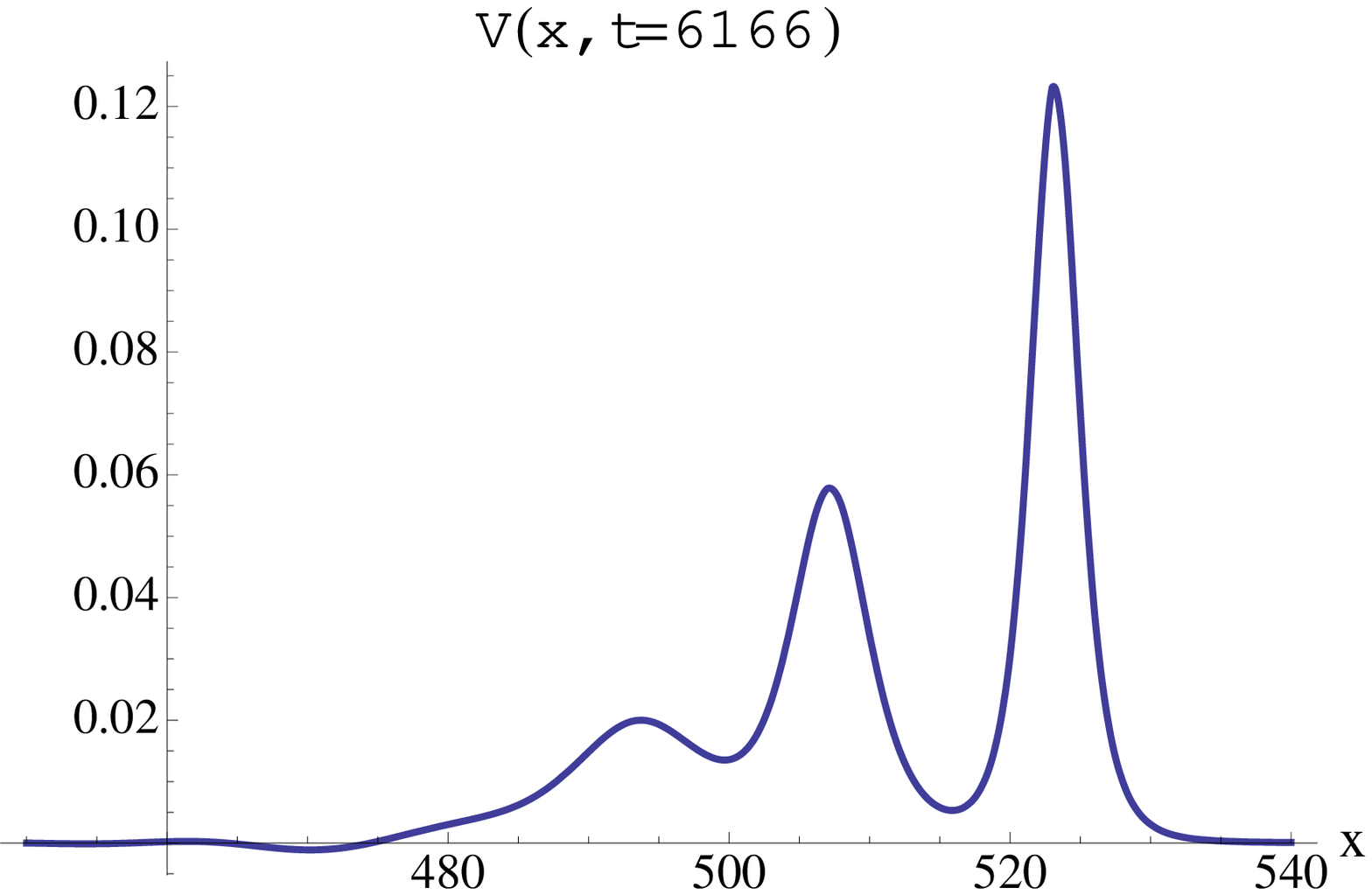}
\caption{\small  
Profile of the velocity field for dipole interaction  $\gamma \gg 1$ (top) and $\gamma \ll 1$ (bottom)}
\label{velocity}
\end{figure*}
%%%%%%%%%%%%%%%%%%%%%%%%%%%%%%%%%%%%%%%%%%%%%

\begin{figure*}[htbp!]
\includegraphics[width=210pt]{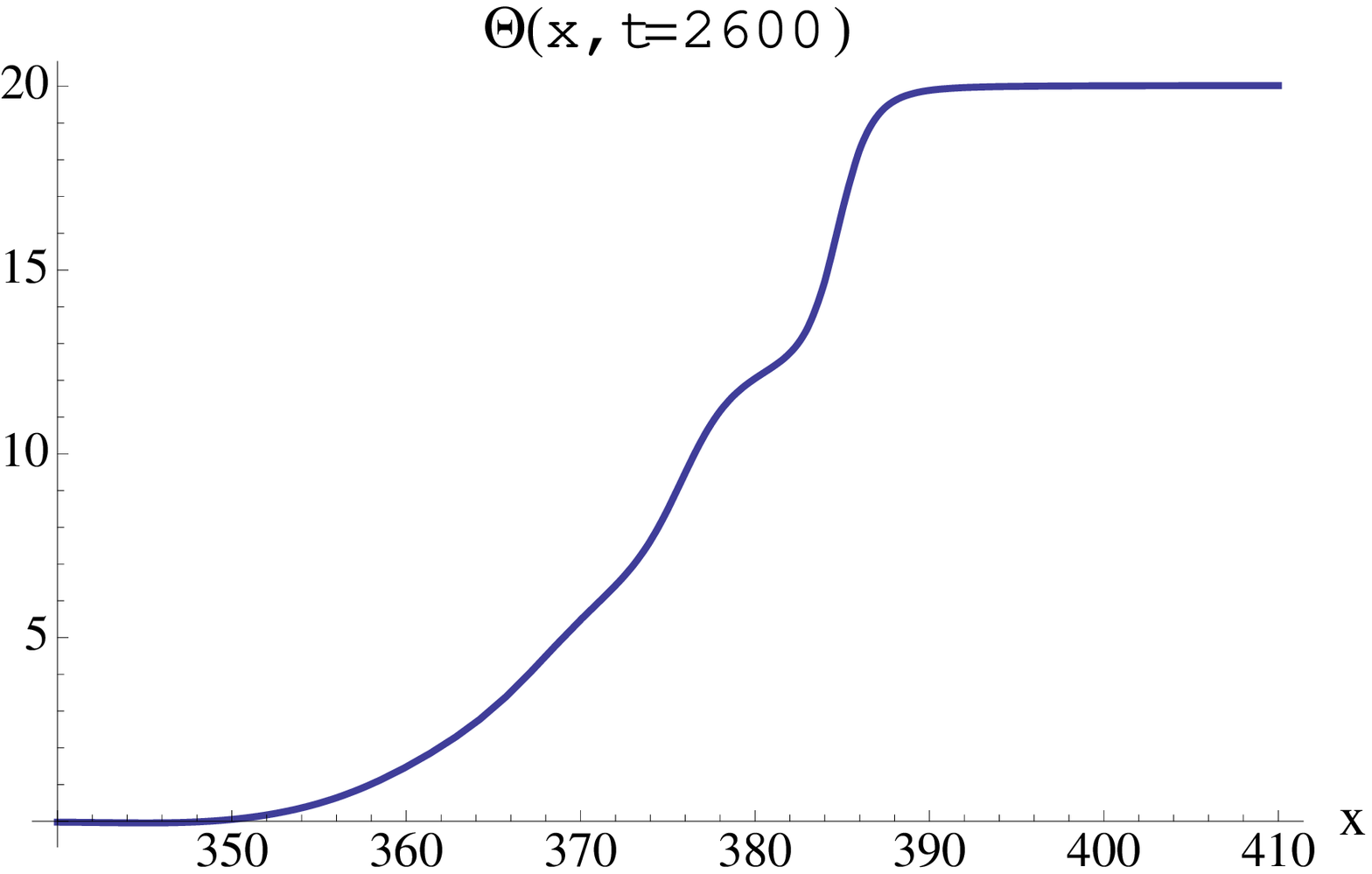}
\includegraphics[width=210pt]{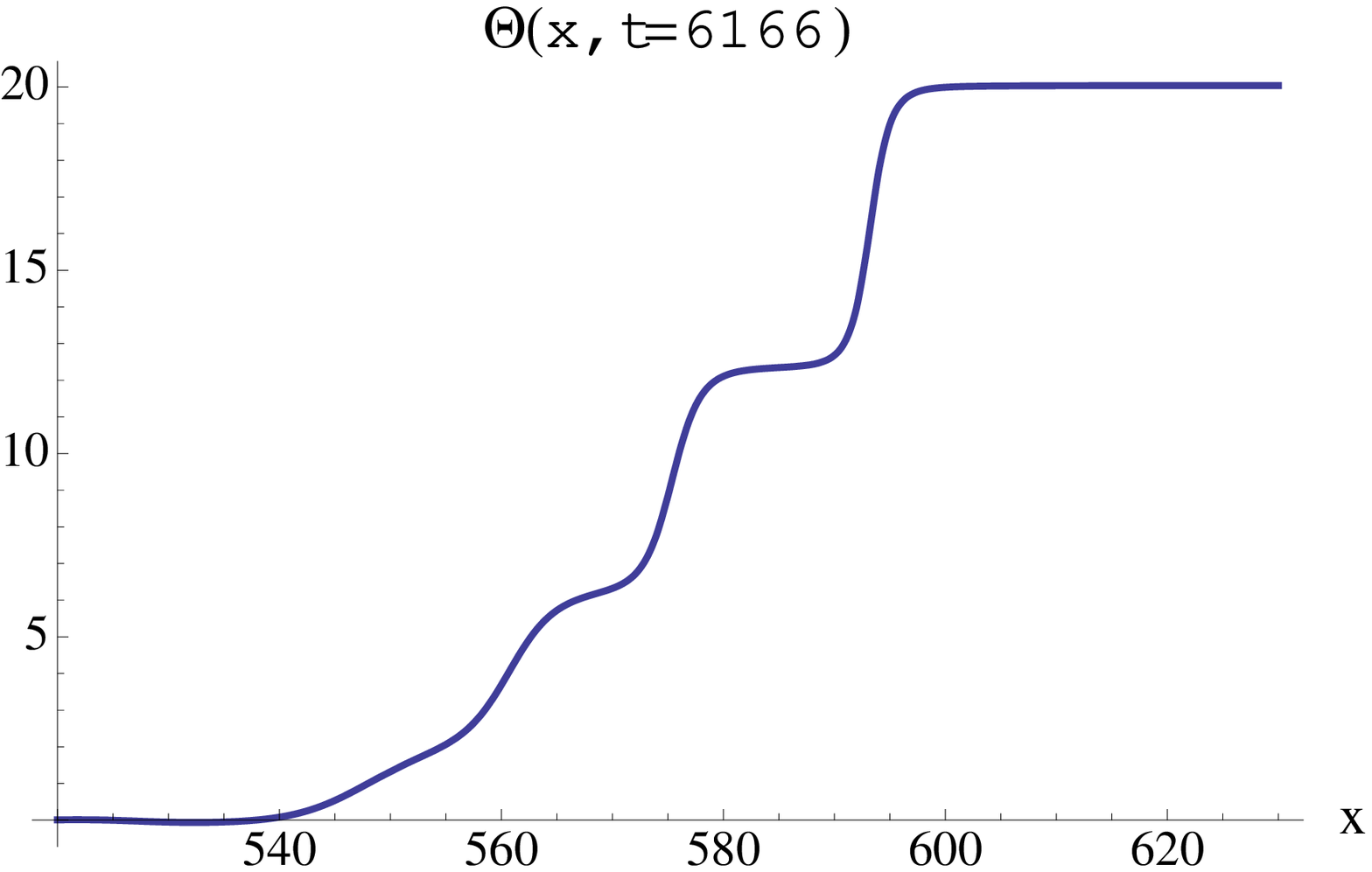}
\caption{\small  
Profile of the $\theta$ field for dipole interaction  $\gamma \gg 1$}
\label{theta}
\end{figure*}
The interference  measurements  probe phase correlation of the Bose-Einstein condensate \cite{Andrews,Shin2004,Schumm,Gati,Berrada,Mutinga}. 
The major  steps of such experiment  are:
(1) a  split of a condensate into two parts, accomplished  by changing of a confining potential  or by  controlling an internal degree of freedom; (2) an independent  evolution of two parts for  a period of time;  (3)  an overlap of two parts. The later creates  a  matter-wave interference pattern. This pattern is usually understood  in terms of the condensate wave function 
\begin{equation}
\label{condensate}
\Psi(x,t)=\sqrt{\rho(x,t)}e^{i\theta(x,t)}\,.
\end{equation}
The interaction plays a multiple-role in an interferometry.   
It influences the dynamics of a condensate, as we have already  seen in   Sections \ref{s2} and \ref{long_range}.
In addition, interaction  also leads to a inleastic scattering  that destroys the quantum coherence.
Note, that the coherence can be also destroyed by another mechanisms, for example on the stage of  when the condensate is split \cite{Altman,Burkov},   
three-body recombinations \cite{Gangardt}, or by an external noise. 
In our analysis we assume that these sources of dephasing are weak, 
and take into account  only the processes that are induced by the density pulse.

\begin{figure*}[htbp!]
\includegraphics[width=210pt]{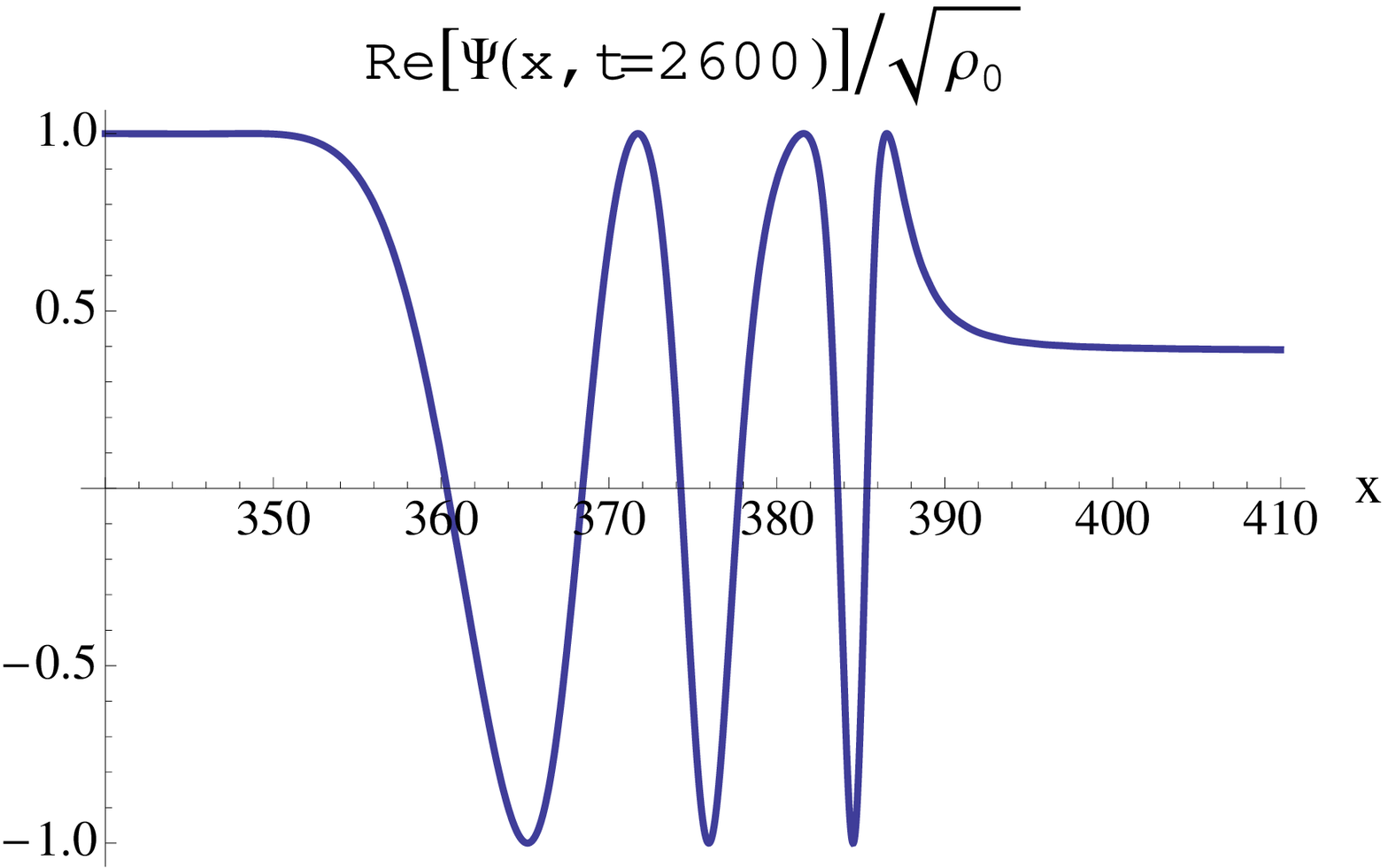}
\includegraphics[width=210pt]{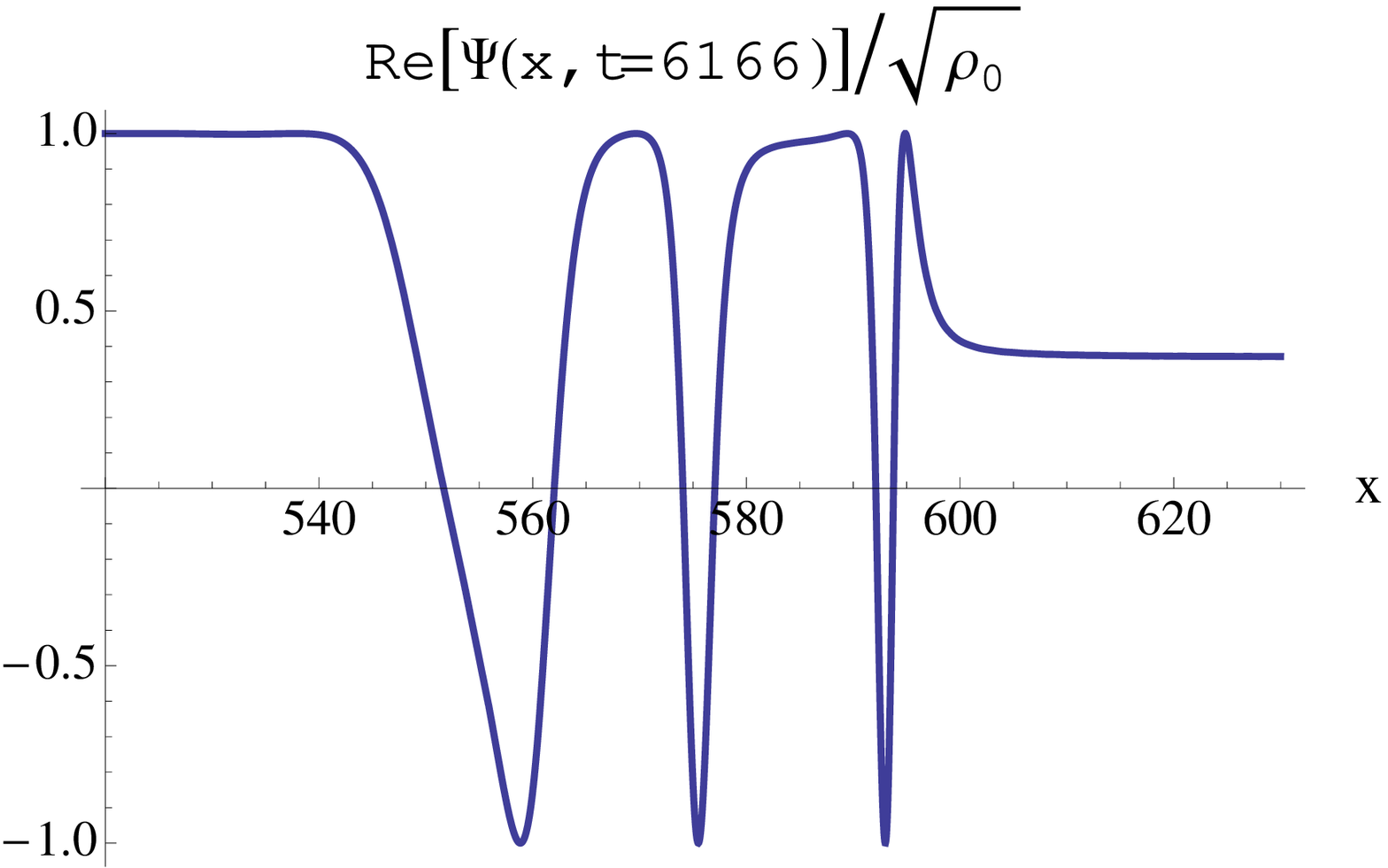}
 \includegraphics[width=210pt]{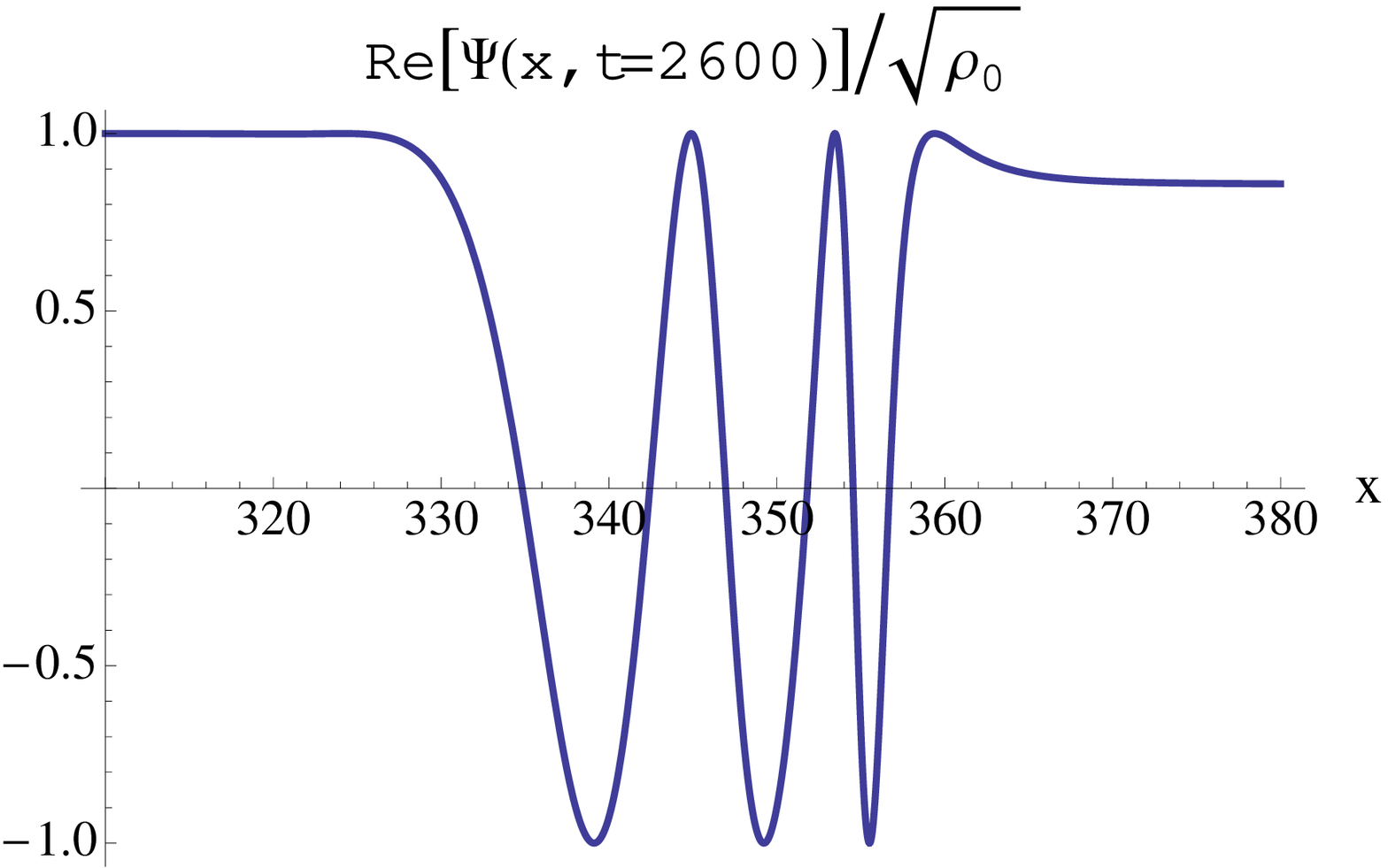}
\includegraphics[width=210pt]{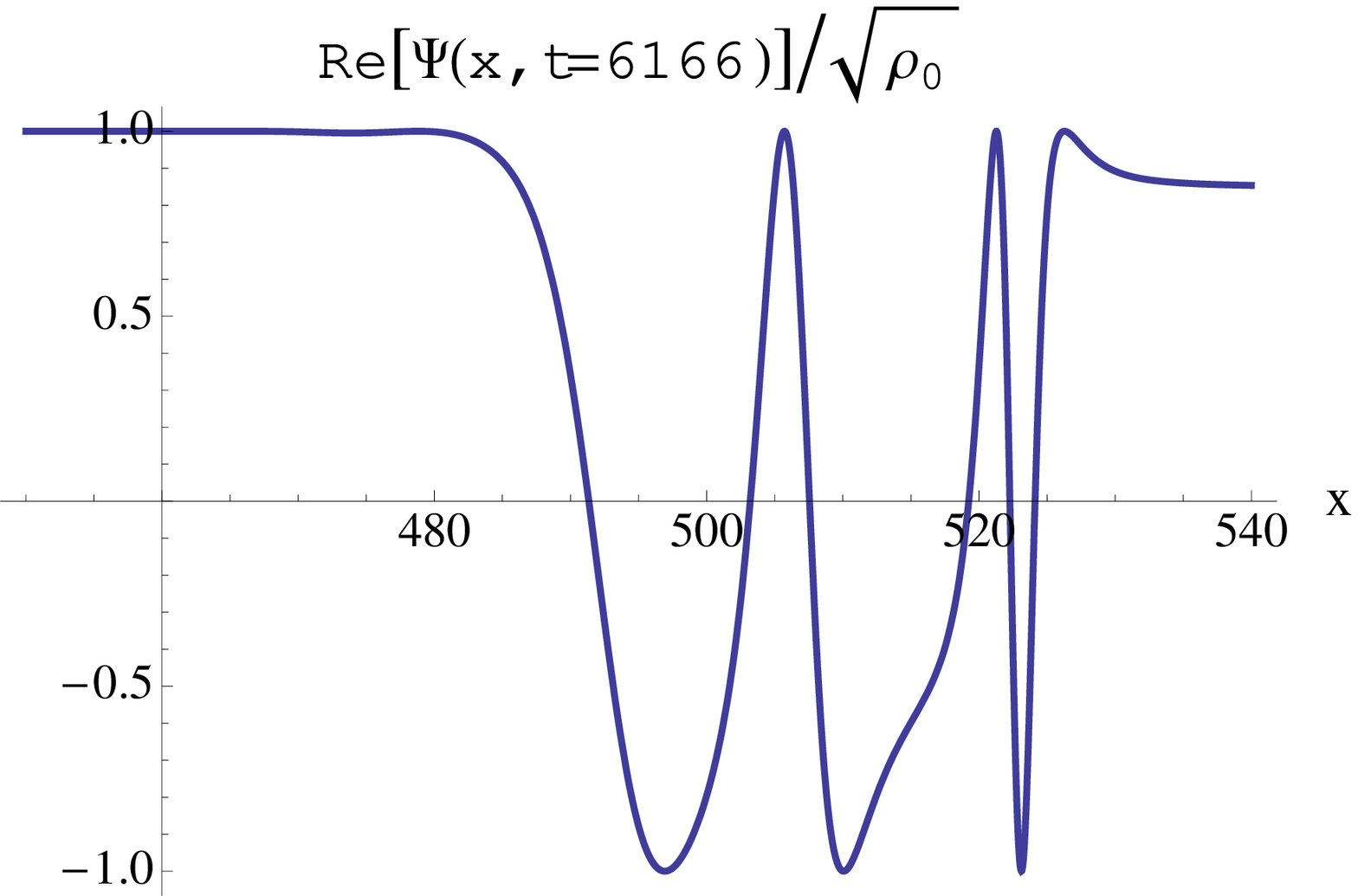}
\caption{\small  
Interference pattern  for  dipole interaction   $\gamma \gg 1$ (top) and $\gamma \ll 1$ (bottom)}
\label{corr}
\end{figure*}

If short range interaction is  strong  ($\gamma \gg 1$) ,  i.e.  in  Tonk-Girardeau limit, 
the bosonic problem can be mapped  onto  the fermionic one.
In this case the  zero temperature inelastic  rate is given by ~\cite{khodas07,Imambekov11}
\begin{equation}
 \frac{1}{\tau_p}\sim
\left[V_0(V_0-V_{k-k_F})\right]^2\frac{(k-k_F)^4}{m^3v_F^6}\,,
\end{equation}
where $V_q$ is the Fourier transform of the  interaction between fermionic particles $V(r)$. 
Estimating the typical  momentum scale $k$ as $k-k_F\sim \Delta \rho$, for interaction  that scales like $1/r^\alpha$
Here, $1\leq \alpha<3$, and the length $l_0$ 
parameterizing the strength of the interaction
is the Bohr radius for the potential $V (r)$.
The inelastic decay rate is thus given by
\begin{equation}
 \frac{1}{\tau_p}\sim \frac{1}{m^7 v_F^6 l_0^8}
(l_0\rho_\infty)^{2\alpha-2}(l_0\Delta\rho)^{2+2\alpha}\,.
\label{ee-decay-rate}
\end{equation}
On the other hand, the characteristic time scale for the density ripples
is given by  Eq. \ref{eq:equation for the shock time strong}.  

Assuming moderate interaction strength $l_0\sim 1/\rho_\infty$ one finds \cite{Protopopov2012}
\begin{equation}
 \frac{t_c}{\tau_p}\sim N\left(\frac{\Delta \rho}{\rho_\infty}\right)^{2\alpha}\,,
\end{equation}
where $N$ is a number of particles contained in the pulse.
We see that in the limit of small $\Delta\rho/\rho_\infty \ll 1$ the characteristic
time $\tau_p$ of inelastic decay  is much larger than the shock time $t_c$.

In the limit of weakly interacting Bose gas ($\gamma \ll 1$)  the life time of bosonic excitations was calculated in Ref.\cite{Matveev2014}.
At zero  temperature, the life time of a long-wave boson is given by
\begin{equation}
\frac{1}{\tau_q} \sim A ^2\frac{\hbar^2\rho_0^2}{m} \left(\frac{q}{q_0} \right)^7\,, \,\,\,{\rm for} \,\,\,q \ll q_0.
\end{equation}
Here $q_0=\sqrt{8mg\rho} $ and $A$  is a  three-body collision amplitude. 
It is equal to zero  in our model, but is  present  in a generic case, for example due to  
excitations of  a  high energy transversal modes\cite{Mazets}.
For large pulse, with $\Delta \rho  \sim q  \gg q_0$  the inelastic length was 
calculated in \cite{Tan}, who found 
\begin{equation}
\frac{1}{\tau} \sim A^2 \frac{\rho^2}{m}\,.
\end{equation}
Since in all limits the  shock formation time $t_c$ remain  finite within integrable Lieb-Liniger model and does 
not depend on $A$ these two time scales are independent. 
Assuming that the integrability is violated only weakly  we will neglect the inelastic processes altogether.

We now analyze the emerging interference pattern of the problem of pulse evolution for the limits 
considered above.  
To cast in a conventional form we use a "condensate" wave function,  Eq.(\ref{condensate}). 
Note, that in one dimension, due to pronounced quantum fluctuation effect the true long range order does not exist.
Nevertheless, one still can formally bosonize the bosonic theory\cite{giamarchi,GGM2011}, 
that results in  the practically identical model.
The bosonic field $\theta$ is related to a velocity field in the usual way
\begin{equation}
v(x,t)=\frac{1}{m}\partial_x\theta(x,t)\,.
\end{equation}
After this transformation both  Euler and continuity equation are  combined into a single non-local Gross-Pitaveskii equation
\begin{eqnarray}&&
\label{Gross_Pitaevskii}
i\partial_t\Psi(x,t)=\left(-\frac{1}{2m}\partial_x^2-\mu+V_{\rm ext}(x,t)\right)\Psi(x,t)  \\&&
+mw_0(|\Psi(x)|^2)\Psi(x)+\int dx'|\Psi(x',t)|^2U(x-x')\Psi(x,t)\,. \nonumber
\end{eqnarray}
Here  we define
\begin{equation}
U(x)=\int \frac{dq}{2\pi}(V(q)-V(0))e^{iqx}\,.
\end{equation}
Thus the solution of non-local Gross-Pitaevskii equation (\ref{Gross_Pitaevskii}) is 
fully equivalent to the non-local hydrodynamic equations we studied earlier.
The profile of the velocity field are shown in Fig.\ref{velocity} for the case of a dipole interaction.
As we wee, the profiles of the  velocity field is quite similar to the density.
This is what we have expected, in our previous estimate. 
Indeed, it follows from the Riemann invariants  for the right (left) moving wave, the functions $\rho$ and $v$
are dependent.   Our numerical simulations show that on the whole, this picture holds in the presence of a generic long range interaction. 
As for the velocity field,  the oscillations for large values of $\gamma$ are more pronounced
than for  small $\gamma$.
The velocity field  $v$ enables us to calculate the field $\theta$, see Fig.\ref{theta}.
As we expected,  picks of the velocity field  correspond to steps in the $\theta$-field.
The later lead to oscillation of the "condesate" wave function 
\begin{equation}
{\rm Re}\left(\Psi(x,t)/\sqrt{\rho(x,t)}\right)=\cos\big(\theta(x,t)\big)\,,
\end{equation}
and the emergence of the interference pattern, shown in Fig. \ref{corr}.
Note, that the phase difference in our case is driven by  the non-equilibrium pulse and arises due to a different length  particles on crests and troughs of the wave have passed. 
%While  parts of the condensate that propagate with higher velocity interfere with parts that propagate with the lower. Because, the velocity and the density fields are interconnected, and oscillate, they give rise
% to an interference pattern.  
By controlling the strength of the initial pulse one is able to change the pattern,
similarly to changing a magnetic field in the conventional Aharonov-Bohm interferometers.

%\begin{figure*}[!ht]
%\centering
%\includegraphics[width=210pt]{C3_Correlation_Low_Gamma_V.png} 
%\includegraphics[width=210pt]{C6_Correlation_Low_Gamma_rho.png} 
%\caption{Velocity  evolution for dipole interaction and $\gamma \ll 1$}
%\label{V_dipole}
%\end{figure*}

%\begin{figure*}[!ht]
%\centering
%\includegraphics[width=210pt]{C3_Correlation_Low_Gamma_theta.png} 
%\includegraphics[width=210pt]{C3_Correlation_Low_Gamma_corr.png} 
%\caption{ $\theta$ field  (on the left) and  $\cos \theta$ (on the right) for $\gamma \ll 1$  and dipole interaction}
%\label{theta_dipole}
%\end{figure*}

\section{Summary and Outlook}
\label{conclusion}

In this paper we  studied  the evolution of a smooth density pulse in a
one dimensional   Bose liquid,  in the presence of   short and  long range  interaction.
At the start of  the evolution the  pulse splits into  a left and a right components, that quickly  separate and propagate independently. 
Prior to the shock formation at $t_c$,  the long range part of interaction plays  a minor role, 
while  short range part of the interaction  determines the enthalpy of the fluid in accordance with 
the exact solution of  Lieb-Liniger model.
By varying the strength of a short range interaction $\gamma$   we were  able to   explore  a variety of 
regimes,  from  Tonk-Girardeau gas  (for $\gamma \gg 1$)  to weakly interacting Bose gas (for 
$\gamma \ll 1$).  

With a formation of a  shock wave,   the long range interaction  starts to  compete with  the non-linear terms, 
stabilizing the solution. The  tug-of-war  between the  nonlinearity and a  long range interaction 
results in  oscillations  of   density and velocity fields.  The scale of these oscillations depends on the type of long range interaction
 and the strength of the short range potential ($\gamma$).

We have computed  the evolution of the density pulse in this regime numerically, 
for the cases of   Coulomb, dipole-dipole  and Van der Waals interaction. 
We  constructed the Riemann invariants for the problem and compared  our  numerical results with analytic estimates.
We found that the   shock formation time ($t_c$), the period of oscillations and the magnitudes of the velocity and density fields  
are in agreement  with our estimates. We found that in all the cases we considered, the interaction induces a regularization that dominates 
over the quantum pressure  term,  in the limit of sufficiently smooth density pulse. 
It implies that the form of the pulse in the long time limit is controlled by the long range part of the interaction, 
even in the limit of Van der Waals interaction.

We have  also studied an  interference pattern of the matter field, induced by a non-equilibrium pulse.
After casting the hydrodynamic equations  in terms of non-local Gross-Pitaevskii
equation, we computed the phase of the condensate wave function.
The oscillations of the velocity field lead to the kinks in  this phase. 
The later manifest itself through interference pattern in time domain.
Our predictions can be tested  experimentally and 
we hope that out work will stimulate such experiments.

\section{Acknowledgments}
We acknowledge useful discussion with A.D. Mirlin and L. Khaykovich.
D.G. acknowledges the support by ISF (grant 584/14),  GIF (grant 1167-165.14/2011),
and Israeli Ministry of Science.

\end{document}